\documentclass[runningheads]{llncs}
\usepackage{graphicx}
\usepackage{tikz}
\usetikzlibrary{automata, positioning, fit}
\usetikzlibrary[positioning,calc,arrows.meta]
\usepackage{amsmath}
\usepackage{amssymb}
\usepackage{hyperref}

\usepackage[capitalise]{cleveref}
\usepackage{makecell}
\usepackage{multicol}
\usepackage{microtype}

\usepackage[utf8]{inputenc}
\usepackage{enumerate}
\usepackage{ifthen}

\usepackage{booktabs}

\makeatletter
\newsavebox{\@brx}
\newcommand{\llangle}[1][]{\savebox{\@brx}{\(\m@th{#1\langle}\)}%
  \mathopen{\copy\@brx\mkern2mu\kern-0.9\wd\@brx\usebox{\@brx}}}
\newcommand{\rrangle}[1][]{\savebox{\@brx}{\(\m@th{#1\rangle}\)}%
  \mathclose{\copy\@brx\mkern2mu\kern-0.9\wd\@brx\usebox{\@brx}}}
\makeatother

\def\Nat{\mathbb{N}}
\def\Real{\mathbb{R}}
\def\Int{\mathbb{Z}}

\DeclareMathOperator{\Prob}{Pr}

\DeclareMathOperator{\Post}{Post}

\renewcommand{\orcidID}[1]{}

\begin{document}
\title{Stochastic Omega-Regular Verification and Control with Supermartingales}

\author{Alessandro Abate\inst{1}\orcidID{0000-0002-5627-9093} 
\and
Mirco Giacobbe\inst{2}\orcidID{0000-0001-8180-0904} 
\and
Diptarko Roy\inst{1}\orcidID{0009-0003-4306-2076}
}

\institute{University of Oxford, UK  \\\email{\{alessandro.abate,diptarko.roy\}@cs.ox.ac.uk}\and
University of Birmingham, UK\\
\email{m.giacobbe@bham.ac.uk}
}

\maketitle              % typeset the header of the contribution
\begin{abstract}
We present for the first time 
a supermartingale certificate 
for $\omega$-regular specifications.  
We leverage the Robbins \& Siegmund convergence theorem to characterize supermartingale certificates for the almost-sure acceptance of Streett conditions on general stochastic processes, which we call Streett supermartingales. This enables effective verification and control of discrete-time stochastic dynamical models with infinite state space under $\omega$-regular and linear temporal logic specifications. 
Our result generalises reachability, safety, reach-avoid, persistence and recurrence specifications; our contribution applies to discrete-time stochastic dynamical models and 
probabilistic programs with discrete and continuous state spaces and distributions, and carries over to deterministic models and programs.
We provide a synthesis algorithm for control policies and Streett supermartingales as proof certificates for $\omega$-regular objectives, which is sound and complete for supermartingales and control policies with polynomial templates and any stochastic dynamical model whose 
post-expectation is expressible as a polynomial. 
We additionally provide an optimisation of our algorithm that reduces
the problem to satisfiability modulo theories, under the assumption that templates and post-expectation are in piecewise linear form. We have built a prototype and have demonstrated the efficacy of our approach on several exemplar $\omega$-regular verification and control synthesis problems.

\keywords{Probabilistic model checking  \and Stochastic control synthesis \and $\omega$-regular verification \and Linear temporal logic \and Martingale theory} 
\end{abstract}

\section{Introduction}

Stochastic processes describe phenomena, systems and computations whose behaviour is probabilistic. 
They are ubiquitous in science and engineering and, in particular, are employed in artificial intelligence and control theory to characterize dynamical models subject to stochastic disturbances, whose correctness is crucial when modelling systems that are deployed to safety-critical environments. Ensuring their correctness with mathematical certainty is an important yet challenging question, in particular for processes with infinite and possibly continuous state spaces. Systems of this kind include sequential decision and planning problems in stochastic environments, auto-regressive time series as well as probabilistic programs, cryptographic protocols, randomised algorithms and much more. Specifications of correctness for complex systems entail complex temporal behaviour, which can be described using linear temporal logic (LTL) or, more generally, $\omega$-regular properties. 

Probabilistic verification algorithms for finite state systems based on explicit-state techniques or symbolic algorithms based on multi-terminal decision diagrams are inapplicable to systems with enumerably infinite or continuous (i.e.\ uncountably infinite) state spaces~\cite{DBLP:conf/cav/KwiatkowskaNP11,DBLP:journals/sttt/HenselJKQV22}. For stochastic processes with infinite state space, existing methods usually build upon finite abstractions or proof rules based on martingale theory.
Finite abstractions first partition the state space into a grid that forms an equivalent (or an approximately equivalent) finite state process, which is then checked using a finite-state verification algorithm~\cite{DBLP:journals/ejcon/AbateKLP10,DBLP:journals/tac/ZamaniEMAL14,bisimulationlearning}. Instead, proof rules directly reduce the verification problem to that of computing proof certificates---known as supermartingale certificates---which are synthesised using constraint solving, guess-and-check procedures, or are learned from data~\cite{DBLP:conf/cav/ChakarovS13,DBLP:conf/cav/ChatterjeeFG16,DBLP:conf/cav/AbateGR20}. 
Proof rules based on supermartingale certificates enable effective verification for infinite-state systems without the intermediate step of computing an abstraction, 
and have been employed with success in the termination and correctness analysis of probabilistic programs as well as the verification of stochastic dynamical models.  

\begin{figure}
\centering
    \begin{tikzpicture}[minimum size=7mm, node distance=14mm]
        \node[draw, circle] (0) {0};
        \node[draw, circle, right of=0] (1) {1};
        \node[draw, circle, right of=1] (2) {2};
        \node[draw, circle, right of=2] (3) {3};
        \node[right of=3] (4) {};
        \node[draw, circle, left of=0] (m1) {-1};
        \node[draw, circle, left of=m1] (m2) {-3};
        \node[draw, circle, left of=m2] (m3) {-5};
        \node[left of=m3] (m4) {};
        \draw (0)  edge[->]node[above] {0.5} (1)
        (1) edge[->] (2) 
        (2) edge[->] (3)
        (3) edge[dashed] (4);
        \draw (0)  edge[->]node[above] {0.5} (m1)
        (m1) edge[->] (m2) 
        (m2) edge[->] (m3)
        (m3) edge[dashed] (m4);

        \draw node[below=0mm of 0] {\color{red} $M$-inc};
        \draw node[below=0mm of 1] {\color{blue} $\epsilon$-dec};
        \draw node[below=0mm of 2] {\color{red} $M$-inc};
        \draw node[below=0mm of 3] {\color{blue} $\epsilon$-dec};
        \draw node[below=0mm of m1] {\color{gray} 0-inc};
        \draw node[below=0mm of m2] {\color{gray} 0-inc};
        \draw node[below=0mm of m3] {\color{gray} 0-inc};

        % \draw node at ($(0)!0.5!(1) - (0,.5)$) {\color{red} $M$-inc};
        % \draw node at ($(1)!0.5!(2) - (0,.5)$) {\color{blue} $\epsilon$-dec};
        % \draw node at ($(2)!0.5!(3) - (0,.5)$) {\color{red} $M$-inc};
        % \draw node at ($(3)!0.5!(4) - (0,.5)$) {\color{blue} $\epsilon$-dec};
        % \draw node at ($(0)!0.5!(m1) - (0,.5)$) {\color{red} $M$-inc};
        % \draw node at ($(m1)!0.5!(m2) - (0,.5)$) {\color{gray} 0-inc};
        % \draw node at ($(m2)!0.5!(m3) - (0,.5)$) {\color{gray} 0-inc};
        % \draw node at ($(m3)!0.5!(m4) - (0,.5)$) {\color{gray} 0-inc};
        
        \draw node[above=0mm of 0] {0};
        \draw node[above=0mm of 1] {1};
        \draw node[above=0mm of 2] {0};
        \draw node[above=0mm of 3] {1};
        \draw node[above=0mm of m1] {1};
        \draw node[above=0mm of m2] {1};
        \draw node[above=0mm of m3] {1};
    \end{tikzpicture}
    \caption{A simple infinite-state stochastic process over variable $x \in \mathbb{Z}$. Above, the value of a Streett supermartingale for the reactivity property ${\sf GF}(x \text{ is even}) \lor {\sf FG} (x < 0)$.}
    \label{fig:simple}
\end{figure}
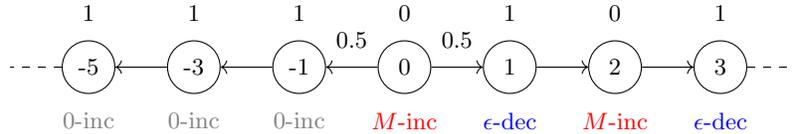
Supermartingale certificates for stochastic models have been developed in the past for specific classes of properties. Previous results introduced proof rules for the almost-sure and the quantitative questions of whether a process eventually hits a target condition (guarantee)~\cite{AgrawalC018,DBLP:conf/cav/ChakarovS13,DBLP:conf/cav/ChatterjeeGMZ22,DBLP:conf/popl/ChatterjeeNZ17}, always avoids an undesirable condition (safety)~\cite{DBLP:conf/sas/ChakarovS14,DBLP:journals/toplas/ChatterjeeFNH18}, and for Boolean combinations of them (obligation), such as reach-avoid specifications~\cite{DBLP:conf/tacas/ChatterjeeHLZ23}. 
Supermartingale certificates were further generalised to 
the properties for which the system eventually satisfies a condition permanently (persistence)~\cite{DBLP:conf/tacas/ChakarovVS16,anand2024compositional}, or hits it infinitely often (recurrence)~\cite{DBLP:conf/tacas/ChakarovVS16}. Yet, arbitrary Boolean combinations of the latter two, which define the $\omega$-regular properties (reactivity) and include LTL~\cite{DBLP:conf/podc/MannaP89}, are beyond reach for existing techniques. 
This includes the example of Fig.~\ref{fig:simple}, which exhibits a process over one integer variable $x$ that, when initialised at $x = 0$, chooses with 0.5 probability to either enumerate the positive numbers or the odd negative numbers. 
This process satisfies almost surely the property requiring that either $x$ is even infinitely often or that $x$ stays strictly negative from some time onwards; however, as we illustrate in Sect.~\ref{sec:related}, previous proof rules cannot verify this property. 

Notably, reducing $\omega$-regular verification to B\"{u}chi acceptance does not easily apply to stochastic processes~\cite{DBLP:journals/apal/Vardi91,DBLP:conf/popl/CookGPRV07,DBLP:conf/hybrid/MuraliTZ24}. 
This is because, to express $\omega$-regular as well as LTL properties, this introduces nondeterminism for which standard martingale theory falls short. To reason about $\omega$-regular specifications while preserving the probabilistic nature of the system, it is necessary to reason about Rabin, Streett, Muller or Parity acceptance conditions, as the respective automata express $\omega$-regular languages in their deterministic form~\cite{Safra1988,DBLP:books/daglib/0020348}. 

We introduce a proof rule for the probabilistic verification of Streett acceptance conditions. Our proof rule leverages the Robbins \& Siegmund convergence theorem for nonnegative almost supermartingales~\cite{Anbar1976,RobbinsSiegmund1971}, which we show to characterise the almost-sure acceptance of Streett pairs. A Streett pair $(A,B)$ is satisfied when either $A$ is visited finitely many times or $B$ is visited infinitely often. To conclude that a stochastic process satisfies a Streett pair $(A,B)$ almost surely, we show that it is sufficient to present a nonnegative real function of the state space that  strictly decreases in expectation when visiting $A\setminus B$,  possibly increases in expectation when visiting $B$, and never increases in expectation in any other case. Such functions---which we call {\em Streett supermartingales}---constitute formal proof certificates that stochastic processes satisfy Streett pairs almost surely. For example, consider the reactivity property in \cref{fig:simple}, which corresponds to the Streett pair where $A = \{ x \mid x \geq 0\}$ and $B = \{ x \mid x \text{ is even}\}$. 
A Streett supermartingale for $(A,B)$ is a function $V \colon \Int \to \Real_{\geq 0}$ that strictly decreases in expectation when visiting positive odd numbers, possibly increases in expectation when visiting nonnegative even numbers, and does not increase in expectation when visiting negative numbers: a valid Streett supermartingale is the function $V(x)$ 
that takes value $1$ if $x$ is odd and takes 
value $0$ if $x$ is even. Notably, for general Streett acceptance conditions with multiple pairs, it suffices to compute one Streett supermartingale for each pair.

Our result enables effective and automated $\omega$-regular 
and LTL verification and control of discrete-time stochastic dynamical models.
We leverage our novel proof rule together with the 
standard result that deterministic Streett automata (DSA) recognise $\omega$-regular languages. 
Our proof rule readily applies to the synchronous 
product between a stochastic process and a DSA, where it suffices to compute one Streett supermartingale for each Streett pair together with a supporting invariant, essential to exclude unreachable states for which the specification fails to hold.
We provide an automated synthesis algorithm to 
compute a (1) Streett supermartingale for each pair, (2) a supporting invariant and (3) a control policy simultaneously, with one call to a decision procedure. 

We show that for time-homogeneous Markov processes with real-valued state space and piecewise polynomial post-expectation, synthesising Streett supermartingales, supporting invariants and policies with piecewise polynomial template of known degree reduces to quantifier elimination over the first-order theory of the reals with one quantifier alternation. Moreover, we show that synthesising piecewise linear Streett supermartingales and policies, 
and polyhedral supporting invariants for processes with piecewise linear post-expectation reduces to the first-order existential theory of reals. 
Finally, we show that when a polyhedral inductive invariant is externally provided, then the synthesis of piecewise linear controllers and Streett supermartingales reduces to quadratically-constrained 
programming (QCP); furthermore, when the system is autonomous, the sole synthesis of Streett supermartingales reduces to linear programming (LP).

We showcase the practical efficacy of our method 
on continuous-state probabilistic systems with piecewise affine dynamics, with a prototype implementation. Our implementation is fully automated  and capable of synthesizing Streett supermartingale certificates, supporting invariants and control policies simultaneously with a single invocation of a satisfiability modulo theory (SMT) solver. As an experimental benchmark, we consider a collection of $\omega$-regular properties ranging over safety, guarantee, recurrence, persistence and reactivity. This demonstrates that our approach is computationally feasible in practice and that it effectively unifies and generalises prior work on supermartingale certificates. 

Our contribution is threefold: 
we present theory, methods, and experiments 
for a novel approach to  
automated stochastic $\omega$-regular verification and control. 

\begin{description}
\item [Theory] We introduce the first supermartingale certificate
for full $\omega$-regular specifications: the Streett supermartingale. By preserving the probabilistic nature of the model, our proof rule enables effective $\omega$-regular verification of infinite state models by reasoning about their post-expectation. 
\item [Methods] We provide sound and complete algorithms for $\omega$-regular verification and control based on our proof rule. Our algorithms compute Streett supermartingales, supporting invariants and control policies with known templates, and are complete relative to provided templates and post-expectations. 
\item [Experiments] We have built a prototype showcasing the efficacy of our algorithms on a set of continuous-state probabilistic systems and $\omega$-regular properties that include and extend beyond the scope of existing approaches. 
\end{description}
Our theoretical contribution applies to any discrete-time deterministic and sto\-chastic dynamical model as well as deterministic and probabilistic programs with discrete and continuous distributions, whose semantics are all special cases of  general stochastic processes. 
Our synthesis algorithm applies to any model whose post-expectation is expressible in piecewise polynomial closed form. 

\section{Streett Supermartingales}\label{sec:streett-sm}

We define stochastic processes on a filtered probability space 
%$(\Omega, {\cal F}, \{{\cal F}_t \}, \Prob)$,
whose space of outcomes $\Omega$ defines an $\cal F$-measurable space of infinite runs, 
and $\{{\cal F}_t \}$ is the associated filtration ${\cal F}_t \subseteq {\cal F}_{t+1} \subseteq {\cal F}$ for all $t \geq 0$.  
A discrete-time stochastic process over a $\Sigma$-measurable state 
space $S$ is a sequence $\{ X_t \}$ with $X_t \colon \Omega \to S$ 
that maps every outcome to the state of a trajectory at time $t$. 
We say that $\{X_t\}$ is adapted to $\{{\cal F}_t \}$ if every $X_t$ is ${\cal F}_t$-measurable, 
namely for all $A \in \Sigma$ it holds that $X_t^{-1}[A] \in {\cal F}_t$. A trajectory $\tau$ is an infinite sequence of states $\tau = \tau_0, \tau_1, \tau_2, \dots$
such that $\tau_t = X_t(\omega)$ for all $t \geq 0$, for some $\omega \in \Omega$. 
Stochastic processes provide a general characterisation for the semantics of stochastic dynamical models described as stochastic difference equations as well as reactive probabilistic programs that run over infinite time. 

Our supermartingale certificate for almost-sure $\omega$-regular verification and control of stochastic processes is underpinned by the Robbins \& Siegmund theorem for the convergence of \textit{nonnegative almost supermartingales}.
\begin{theorem}[{\protect Robbins \& Siegmund Convergence Theorem \cite{RobbinsSiegmund1971}}]\label{thm:robbins-siegmund}
Let $\{{\cal F}_t \}$ be a filtration and let 
$\{V_t \}$, 
$\{U_t \}$, and
$\{W_t \}$ 
be three real-valued nonnegative stochastic processes adapted to $\{ {\cal F}_t \}$.
Suppose that, for all $t \in \Nat$, the following statement holds almost surely:
\begin{align}
    E ({V_{t+1}} \mid { \mathcal{F}_t }) &\leq V_t-U_t+W_t. 
\label{eqn:drift-requirement-smconv-thm}
\end{align}
Then,
\begin{equation}
    \Prob\left(  \sum_{t = 0}^\infty U_t < \infty \lor \sum_{t = 0}^\infty W_t = \infty \right) = 1.
\end{equation}
\end{theorem}
This result generalises the classic convergence theorem for nonnegative supermartingales~\cite[Theorem 22, p.148]{pollard_2001}, allowing the real-valued process $\{ V_t \}$ to satisfy the weaker almost-supermartingale condition of \cref{eqn:drift-requirement-smconv-thm} with respect to the two other real-valued processes $\{ U_t \}$ and $\{ W_t \}$~\cite{Anbar1976,RobbinsSiegmund1971}. 
The statement establishes that the event that 
either series $\sum_{t = 0}^\infty U_t$ converges or series
$\sum_{t = 0}^\infty W_t$ diverges has probability 1. As we show below, this naturally characterises almost-sure Streett acceptance for general stochastic processes. 

A Streett pair $(A,B)$ consists of two measurable regions of 
the state space $A,B \in \Sigma$. 
A trajectory $\tau = \tau_0, \tau_1, \tau_2, \dots$ satisfies $(A,B)$
if either it visits all states in $A$ finitely many times or 
it visits any states in $B$ infinitely many times; more formally, 
$\tau$ satisfies $(A,B)$ if 
$\sum_{t = 0}^\infty {\bf 1}_{A_i}(\tau_t) < \infty \lor \sum_{t = 0}^\infty {\bf 1}_{B_i}(\tau_t) = \infty
$, 
where ${\bf 1}_{\cal S}(\cdot)$ denotes the indicator function of set $\cal S$, which takes value 1 when its argument is a member of $\cal S$ and takes value 0 otherwise. Our result establishes that, to conclude that a stochastic process $\{ X_t \}$ 
satisfies $(A,B)$ almost surely, it suffices to present a 
function $V$ that maps $\{ X_t \}$ to a nonnegative 
almost-supermartingale whose expected value decreases strictly when visiting $A\setminus B$, possibly increases when visiting $B$, and never increases anywhere else almost surely. We call function $V$ a Streett supermartingale for $(A,B)$.

\begin{theorem}[Streett Supermartingales]\label{thm:streett-supermartingale}
Let $\{ X_t\}$ be a stochastic process over state space $S$ and $(A, B)$ be a Streett pair.  
Suppose that
there exists a nonnegative function $V : S \to \Real_{\geq 0}$ and positive constants $\epsilon, M > 0$ such that, for all $t \in \Nat$, the following condition holds almost surely:
\begin{align}
    E[
    V(X_{t+1}) \mid {\cal F}_t
    ]
    \leq V(X_t) - \epsilon \cdot \mathbf{1}_{A \smallsetminus B}(X_t) + M\cdot \mathbf{1}_B(X_t).\label{eq:streett-supermartingale}
\end{align}
Then, $\{ X_t \}$ satisfies $(A,B)$ almost surely, i.e., 
\begin{equation}
    \Prob\left( \sum_{t = 0}^\infty {\bf 1}_A(X_t) < \infty \lor   \sum_{t = 0}^\infty {\bf 1}_B(X_t) = \infty \right) = 1.\label{eqn:conclusion-thm2}
\end{equation}
%We call $V$ a Streett supermartingale for $(A,B)$.
\end{theorem}

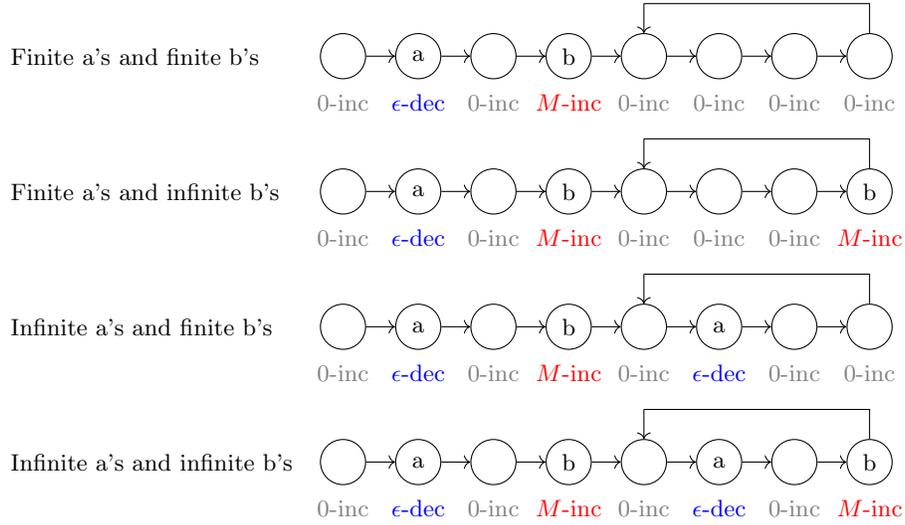
\begin{figure}
\centering
    \begin{tikzpicture}[minimum size=6mm]
        \def\vdist{18mm}

        \node (m1) {Finite a's and finite b's};
        \node[draw, circle] at ($(m1.east)+(10mm,0)$)(m10) {};
        \node[draw, circle, right of=m10] (1) {a};
        \node[draw, circle, right of=1] (2) { };
        \node[draw, circle, right of=2] (3) {b};
        \node[draw, circle, right of=3] (4) { };
        \node[draw, circle, right of=4] (5) { };
        \node[draw, circle, right of=5] (6) { };
        \node[draw, circle, right of=6] (7) { };
        \draw (m10) edge[->] (1) 
        (1) edge[->] (2) 
        (2) edge[->] (3) 
        (3) edge[->] (4) 
        (4) edge[->] (5) 
        (5) edge[->] (6) 
        (6) edge[->] (7) 
        (7) -- ($(7)+(0,.7)$) -- ($(4)+(0,.7)$) edge[->] (4);
        \draw node[below=0mm of m10] {\color{gray} $0$-inc};
        \draw node[below=0mm of 1] {\color{blue} $\epsilon$-dec};
        \draw node[below=0mm of 2] {\color{gray} $0$-inc};
        \draw node[below=0mm of 3] {\color{red} $M$-inc};
        \draw node[below=0mm of 4] {\color{gray} $0$-inc};
        \draw node[below=0mm of 5] {\color{gray} $0$-inc};
        \draw node[below=0mm of 6] {\color{gray} $0$-inc};
        \draw node[below=0mm of 7] {\color{gray} $0$-inc};
        
        \node[below= \vdist of $(m1.west)$, anchor=west] (m2) {Finite a's and infinite b's};
        \node[below= \vdist of $(m10.west)$, anchor=west, draw, circle] (m20) {};
        \node[draw, circle, right of=m20] (1) {a};
        \node[draw, circle, right of=1] (2) { };
        \node[draw, circle, right of=2] (3) {b};
        \node[draw, circle, right of=3] (4) { };
        \node[draw, circle, right of=4] (5) {};
        \node[draw, circle, right of=5] (6) { };
        \node[draw, circle, right of=6] (7) {b};
        \draw (m20) edge[->] (1) 
        (1) edge[->] (2) 
        (2) edge[->] (3) 
        (3) edge[->] (4) 
        (4) edge[->] (5) 
        (5) edge[->] (6) 
        (6) edge[->] (7) 
        (7) -- ($(7)+(0,.7)$) -- ($(4)+(0,.7)$) edge[->] (4);
        \draw node[below=0mm of m20] {\color{gray} $0$-inc};
        \draw node[below=0mm of 1] {\color{blue} $\epsilon$-dec};
        \draw node[below=0mm of 2] {\color{gray} $0$-inc};
        \draw node[below=0mm of 3] {\color{red} $M$-inc};
        \draw node[below=0mm of 4] {\color{gray} $0$-inc};
        \draw node[below=0mm of 5] {\color{gray} $0$-inc};
        \draw node[below=0mm of 6] {\color{gray} $0$-inc};
        \draw node[below=0mm of 7] {\color{red} $M$-inc};
        
        \node[below= \vdist of $(m2.west)$, anchor=west]  (m3) {Infinite a's and finite b's};
        \node[draw, circle, below= \vdist of $(m20.west)$, anchor=west] (m30) {};
        \node[draw, circle, right of=m30] (1) {a};
        \node[draw, circle, right of=1] (2) { };
        \node[draw, circle, right of=2] (3) {b};
        \node[draw, circle, right of=3] (4) { };
        \node[draw, circle, right of=4] (5) {a};
        \node[draw, circle, right of=5] (6) { };
        \node[draw, circle, right of=6] (7) { };
        \draw (m30) edge[->] (1) 
        (1) edge[->] (2) 
        (2) edge[->] (3) 
        (3) edge[->] (4) 
        (4) edge[->] (5) 
        (5) edge[->] (6) 
        (6) edge[->] (7) 
        (7) -- ($(7)+(0,.7)$) -- ($(4)+(0,.7)$) edge[->] (4);
        \draw node[below=0mm of m30] {\color{gray} $0$-inc};
        \draw node[below=0mm of 1] {\color{blue} $\epsilon$-dec};
        \draw node[below=0mm of 2] {\color{gray} $0$-inc};
        \draw node[below=0mm of 3] {\color{red} $M$-inc};
        \draw node[below=0mm of 4] {\color{gray} $0$-inc};
        \draw node[below=0mm of 5] {\color{blue} $\epsilon$-dec};
        \draw node[below=0mm of 6] {\color{gray} $0$-inc};
        \draw node[below=0mm of 7] {\color{gray} $0$-inc};
        
        \node[below= \vdist of $(m3.west)$, anchor=west]  (m4) {Infinite a's and infinite b's};
        \node[draw, circle, below= \vdist of $(m30.west)$, anchor=west] (m40) {};
        \node[draw, circle, right of=m40] (1) {a};
        \node[draw, circle, right of=1] (2) { };
        \node[draw, circle, right of=2] (3) {b};
        \node[draw, circle, right of=3] (4) { };
        \node[draw, circle, right of=4] (5) {a};
        \node[draw, circle, right of=5] (6) { };
        \node[draw, circle, right of=6] (7) {b};
        \draw (m40) edge[->] (1) 
        (1) edge[->] (2) 
        (2) edge[->] (3) 
        (3) edge[->] (4) 
        (4) edge[->] (5) 
        (5) edge[->] (6) 
        (6) edge[->] (7) 
        (7) -- ($(7)+(0,.7)$) -- ($(4)+(0,.7)$) edge[->] (4);
        \draw node[below=0mm of m40] {\color{gray} $0$-inc};
        \draw node[below=0mm of 1] {\color{blue} $\epsilon$-dec};
        \draw node[below=0mm of 2] {\color{gray} $0$-inc};
        \draw node[below=0mm of 3] {\color{red} $M$-inc};
        \draw node[below=0mm of 4] {\color{gray} $0$-inc};
        \draw node[below=0mm of 5] {\color{blue} $\epsilon$-dec};
        \draw node[below=0mm of 6] {\color{gray} $0$-inc};
        \draw node[below=0mm of 7] {\color{red} $M$-inc};
        
    \end{tikzpicture}
    \caption{Intuition for \Cref{thm:streett-supermartingale} on exemplar trajectories.}
    \label{fig:intuition}
\end{figure}
\begin{example}\label{example:simple-reactivity}
\Cref{fig:intuition} illustrates \cref{thm:streett-supermartingale} over four exemplar trajectories, 
with respect to the Streett pair $(\{ s \mid s \text{ has label a}\}, \{ s \mid s \text{ has label b}\})$. 
In this example, we illustrate that 
a Streett supermartingale $V$---which must be nonnegative---cannot be constructed for the third trajectory, as \cref{eq:streett-supermartingale} requires $V$ to strictly decrease by $\epsilon$ infinitely many times in the tail behaviour of the trajectory while being never allowed to increase. For all other trajectories instead, a Streett supermartingale $V$ and suitable constants  $\epsilon, M > 0$ exist. In particular, in the first and second trajectories any $V$ is only required to strictly decrease finitely many times. In the fourth trajectory, $V$ is permitted to compensate its requirement to 
decrease infinitely many times by increasing infinitely many times in the tail behaviour. Notably, the first, the second, and the fourth trajectory are precisely those trajectories that satisfy the specification. \qed 
\end{example}

We provide a specialisation of \Cref{thm:streett-supermartingale} (which applies to general stochastic processes) to time-homogeneous Markov processes, whose dynamics only depend on their transition kernel.
A {\em transition kernel} $T \colon S \times \Sigma \to [0,1]$ gives the probability that the process makes a transition from state $s \in S$ into the set $S^\prime \in \Sigma$, independently of time, i.e., for all $t \in \Nat$, $T(X_t, S') = \Prob(X_{t+1} \in S' \mid {\cal F}_t)$.
The transition kernel in turn determines the {\em post-expectation} $(\Post h) \colon S \to \Real$ of any real-valued measurable function $h \colon S \to \Real$,
defined as the conditional expectation of $h$ after one time step 
(regardless of absolute time $t$) as follows:
\begin{equation}\label{eqn:postex}
    \Post h(X_t) = \int_S
    h(s)~T(X_t, \mathrm{d}s)
    = E(h(X_{t+1}) \mid {\cal F}_t). 
\end{equation}
This denotes the expected value of $h$ 
when evaluated in the subsequent state, given the current state 
being $X_t$. 
For time-homogeneous Markov processes, we establish that to obtain a valid 
Streett supermartingale it suffices to enforce the requirement 
of \cref{eq:streett-supermartingale} over $\Post V$ of a Streett supermartingale $V$ whose domain is restricted to a sufficiently strong supporting invariant $I$.
\begin{theorem}[Supporting Invariants]
\label{thm:requirements-corollary}
Let $\{ X_t \}$ be a time-homogeneous Mar\-kov process with initial state $s_0 \in S$ and transition kernel $T \colon S \times \Sigma \to [0,1]$. Let $(A,B)$ be a Streett pair. Suppose there exists a measurable set $I \in \Sigma$, a nonnegative function $V \colon I \to \Real_{\geq 0}$ and positive constants $\epsilon,M > 0$ that satisfy the following five conditions:
\begin{align}
&s_0 \in I\label{eq:init}\\
&\forall s \in I \colon T(s, I) = 1\label{eq:invar}\\
&\forall s \in (A \setminus B) \cap I \colon 
    \Post V(s) \leq V(s) - \epsilon\label{eq:dec}\\
&\forall s \in B \cap I \colon
    \Post V(s) \leq V(s) + M\label{eq:inc}\\
&\forall s \in I \setminus (A \cup B) \colon 
    \Post V(s) \leq V(s)\label{eq:noninc}
\end{align}
Then, $V$ is a Streett supermartingale for $(A,B)$.
\end{theorem}

\begin{example} Consider the time-homogeneous Markov process in \Cref{fig:simple} and the 
LTL property ${\sf GF}(x \text{ is even})$, corresponding to the Streett pair $(\Int, \{ x \mid x \text{ is even}\})$. 
Provided the supporting invariant $\{ x \in \Int \mid x > 0\}$, the function that maps the positive even numbers to 0 and the positive odd numbers to 1 is a valid 
Streett supermartingale if the process is initialised on a positive number. Without a supporting invariant, function $V$ would be required to strictly decrease along all negative numbers, necessarily violating nonnegativity. Notably, the process satisfies ${\sf GF}(x \text{ is even})$ almost surely only on the positive numbers. \qed
\end{example}

Finally, a general Streett acceptance condition consists of a 
finite set of Streett pairs, and a trajectory satisfies the acceptance condition if it satisfies all pairs. To establish that a stochastic process satisfies a general Streett acceptance condition, it suffices to present one Streett supermartingale for each pair. 
\begin{theorem}\label{thm:general-streett-pairs}
Let $\{ X_t \}$ be a stochastic process and 
%Furthermore, let 
$\{ (A_i, B_i) \colon i = 1, \ldots, k\}$ be a Streett acceptance condition. If every Streett pair admits a Streett supermartingale, then $\{ X_i \}$ satisfies the acceptance condition almost surely:
\begin{equation}
    \Prob\left( \bigwedge_{i=1}^k \left ( \sum_{t = 0}^\infty {\bf 1}_{A_i}(X_t) < \infty \lor \sum_{t = 0}^\infty {\bf 1}_{B_i}(X_t) = \infty \right)\right) = 1. 
\end{equation}
\end{theorem}

\section{Stochastic Omega-regular Verification and Control}\label{sec:verification}
A stochastic dynamical model $\cal M$ over $\Real^n$ consists of an initial state vector $x_0 \in \Real^n$ and a parameterised update function $f \colon \Real^n \times {\cal W} \times K \to \Real^n$ with a space ${\cal W} $ of input disturbances and 
a space $K$ of control parameters. 
This defines a time-homogeneous Markov process over 
the ${\cal B}(\Real^n)$-measurable state space $\Real^n$ given by the following equation: 
\begin{equation}\label{eqn:update-func}
    X_{t+1}^{\cal M} = f(X_t^{\cal M}, W_t; \kappa), \quad X_0^{\cal M} = x_0,
\end{equation}
where $\{ W_t \}$ is a sequence of i.i.d.\ stochastic input disturbances, 
each of which draws from the sample space $\cal W$. This assumption restricts 
our model to time-homogeneous Markov processes, 
for which \Cref{thm:requirements-corollary} applies.
This model subsumes autonomous systems as well as control systems with 
parameterised policies. For example, a stochastic dynamical model 
$f' \colon \Real^n \times {\cal U} \times {\cal W} \to \Real^n$  
with finite or infinite space of control inputs $\cal U$ and a parameterised 
(memoryless deterministic) policy $\pi \colon \Real^n \times K \to \cal U$ results in the 
special case $f(x, w; \kappa) = f'(x, \pi(x; \kappa), w)$. Notably, our model
also encompasses finite memory policies 
with known template and known memory size, for which it is sufficient to 
add extra state variables and extra input disturbances. 

We associate our model with a finite set of observable propositions $\Pi$ and an observation function $\llangle \cdot \rrangle \colon \Real^n \to 2^\Pi$ that maps every state to the set of propositions that hold true in that state. This defines a (measurable) set of traces---the trace language of $\cal M$---where a trace $\hat\tau$ is an infinite sequence 
${\hat\tau} = {\hat\tau}_0,{\hat\tau}_1,{\hat\tau}_2, \dots$
where ${\hat\tau}_i = \llangle \tau_i \rrangle$ for all $i \geq 0$,
with $\tau = \tau_0, \tau_1, \tau_2, \dots$ being some trajectory of $\{ X_t^{\cal M} \}$. We treat the question of synthesizing a controller for which 
$\cal M$ satisfies an LTL formula over atomic propositions in $\Pi$ or, 
more generally, satisfies an $\omega$-regular property over alphabet $2^\Pi$ almost surely.
For this purpose, we leverage the standard result that deterministic Streett automata (DSA) 
recognise the $\omega$-regular languages. The control synthesis problem amounts 
to computing a control parameter $\kappa \in K$ for which the event that the trace 
language of $\cal M$ is accepted by DSA $\cal A$ has probability 1. 
The verification problem for autonomous systems or systems with fixed control policy
can be simply seen as the special case where $K$ is a singleton.  

A deterministic Streett automaton $\cal A$ over alphabet $2^\Pi$ consists of a finite set of states $Q$, an initial state $q_0$, a transition function $\delta \colon Q \times 2^\Pi \to Q$, 
and a finite set of Streett pairs $\mbox{Acc} = \{(A_1, B_1), \dots, (A_k,B_k)\}$ where $A_i \subseteq Q$ and $B_i \subseteq Q$ for all $i = 1, \dots, k$. The run $\rho$ of $\cal A$ on input trace $\hat\tau = {\hat\tau}_0,{\hat\tau}_1,{\hat\tau}_2, \dots$ is the infinite sequence of states $\rho = \rho_0, \rho_1, \rho_2, \dots$ such that $\rho_0 = q_0$ and $\rho_{t+1} = \delta(\rho_t, \hat\tau_t)$ for all $t \geq 0$. The automaton accepts $\hat\tau$
if either $\rho$ visits $A_i$ finitely many times or $B_i$ infinitely many times for all $i = 1, \dots k$, i.e., $\bigwedge_{i=1}^k \sum_{t=1}^\infty {\bf 1}_{A_i}(\rho_t) < \infty \lor \sum_{t=1}^\infty {\bf 1}_{B_i}(\rho_t) = \infty$.
Our approach to probabilistic $\omega$-regular verification 
leverages the fact that a DSA (indeed, any deterministic automaton) recognising the traces of a stochastic process forms in its turn a stochastic  process:
\begin{equation}
    X_{t+1}^{\cal A} = \delta(X_t^{\cal A}, \llangle X_{t}^{\cal M} \rrangle), \quad X_0^{\cal A} = q_0.
\end{equation}

Our approach determines whether 
$\{ X_{t}^{\cal A} \}$ satisfies the Streett acceptance condition of $\cal A$ with probability 1.
We note that Streett automata are dual to Rabin automata,
thus any tool to translate an LTL formula $\varphi$ 
to a Rabin automaton, equivalently produces a Streett automaton for $\lnot \varphi$~\cite{DBLP:conf/cav/KretinskyMSZ18,DBLP:conf/cav/Duret-LutzRCRAS22}. Our output can thus be equivalently cast as 
Rabin acceptance with probability 0. 

To determine whether $\{X^{\cal A}_t\}$ satisfies the acceptance condition 
of $\cal A$, we leverage \Cref{thm:requirements-corollary,thm:general-streett-pairs} to synthesize a Streett supermartingale for each Streett pair and a supporting invariant over the {\em synchronous product} of $\cal M$ and $\cal A$. This is because the process 
$\{ X_t^{\cal A}\}$ is not time-homogeneous when considered in isolation, as the distribution of next states in the automaton requires 
information about  $\{ X_t^{\cal M}\}$ to be determined.  
Therefore, we define the product process $\{ X_{{t} }^{{\cal M}\otimes{\cal A}} \}$ as $X_{t}^{{\cal M}\otimes{\cal A}} =  (X_{t}^{\cal M}, X_{t}^{\cal A})$ for all $t \in \Nat$, where 
$X_{t+1}^{{\cal M}\otimes{\cal A}} = (f(X^{\cal M}_t, W_t; \kappa), \allowbreak \delta(X^{\cal A}_t, \llangle X^{\cal M}_t \rrangle))$ and 
$X_{0}^{{\cal M}\otimes{\cal A}} = (x_0, q_0)$.
We then extend the acceptance condition of $\cal A$ to 
the product state space $\Real^n \times Q$. Concretely, 
we define $\overline{A}_i = \Real^n \times A_i$ and $\overline{B}_i = \Real^n \times B_i$ for $i = 1,\ldots, k$, and we 
define the acceptance condition of the product process as
$\{ (\overline{A}_1, \overline{B}_1), \ldots, (\overline{A}_k, \overline{B}_k) \}$. Finally, we establish that $\{ X_{t}^{{\cal M}\otimes{\cal A}} \}$ satisfies the given acceptance condition almost surely by computing $k$ Streett supermartingales 
and one supporting invariant over $\Real^n \times Q$. 

We assume a known parameterised form for Streett supermartingales and invariant 
as well as for the control policy (as described above) and, by using 
\Cref{thm:requirements-corollary,thm:general-streett-pairs}, we express the verification and control problem 
as the problem of deciding a quantified 
first-order logic formula. 
Let $V \colon \Real^n \times Q \times \Theta \to \Real_{\geq 0}$ be a parameterised non-negative function of $\Real^n \times Q$ (the Streett supermartingale certificate), 
with parameter space $\Theta$.
The post-expectation of $V$ results in the parameterised function
$(\Post V) \colon \Real^n \times Q \times \Theta \times K \to \Real_{\geq 0}$ 
over the certificate parameters $\Theta$ of $V$ and the control parameters $K$ defined as
\begin{equation}
    \Post V(x,q;\theta,\kappa) = \int_{\cal W} V(f(x,w;\kappa),\delta(q,\llangle x \rrangle); \theta) \Pr(\mathrm{d}w)
\end{equation}
To construct our first-order logic decision problem, it is essential to 
express $\Post V$ in a symbolic closed-form representation. Notably, 
computing symbolic closed-form representations for the post-expectation is a general problem in probabilistic verification, for which automated tools exist~\cite{gehr2016psi}. 
Provided that $\Post V$
is computable, we template $k$ parameterised Streett supermartingale certificates $V_1, \dots, V_k$ 
with parameter spaces $\Theta_1, \dots, \Theta_k$ respectively, and template  
one parameterised invariant predicate $I \colon \Real^n \times Q \times H \to \{ \mathrm{true}, \mathrm{false}\}$ with parameter space $H$. 
Then, solving the $\omega$-regular control problem with our method amounts to searching for certificate parameters $\theta_1 \in \Theta_1, \dots, \theta_k \in \Theta_k$, invariant parameter $\eta \in H$, control parameter $\kappa \in K$ and 
coefficients $\epsilon, M > 0$ such that, 
for every $i = 1, \dots, k$, the following universally quantified sentences hold:
\begin{align}
    &I(x_0, q_0; \eta) \label{eqn:VerifCondition-init}\\
    &\forall x \in \Real^n, q \in Q, w \in \mathcal{W} \colon 
     I(x, q; \eta) \implies 
     I(f(x, w; \kappa), \delta(q, \llangle x \rrangle); \eta)
     \label{eqn:VerifCondition-consecution}\allowdisplaybreaks\\
    &\forall x \in \Real^n, q \in (A_i \setminus B_i) \colon I(x,q; \eta) \implies
     \Post {V_i} (x, q; \theta_i, \kappa) \leq V_i(x, q; \theta_i) - \epsilon \label{eqn:VerifCondition-epsDec} \allowdisplaybreaks\\
    &\forall x \in \Real^n, q \in B_i \colon I(x,q; \eta) \implies
     \Post {V_i} (x, q; \theta_i, \kappa) \leq V_i(x, q; \theta_i) + M
     \label{eqn:VerifCondition-Minc}\allowdisplaybreaks\\
     &\forall x \in \Real^n, q \in Q\setminus(A_i \cup B_i) \colon I(x,q; \eta)\Rightarrow
     \Post {V_i}(x, q; \theta_i, \kappa) \leq V_i(x, q; \theta_i)
     \label{eqn:VerifCondition-0inc}\\
     &\forall x \in \Real^n, q \in Q \colon 
     I(x, q; \eta) \implies V_i(x, q; \theta_i) \geq 0 \label{eqn:VerifCondition-NonNeg}
\end{align}
In particular, \cref{eqn:VerifCondition-init,eqn:VerifCondition-consecution} respectively indicate the conditions of 
\textit{initiation} and \textit{consecution} for the supporting invariant, yielding a subset of the product space satisfying \cref{eq:init,eq:invar}. 
\Cref{eqn:VerifCondition-epsDec,eqn:VerifCondition-Minc,eqn:VerifCondition-0inc} indicate the {\em drift conditions}, which ensure that $V_1, \dots, V_k$ 
satisfy \cref{eq:dec,eq:inc,eq:noninc} w.r.t. the  
acceptance conditions extended to the product space. \Cref{eqn:VerifCondition-NonNeg} enforces the premise of \Cref{thm:requirements-corollary} 
that requires $V$ to be non-negative over its domain $I$.
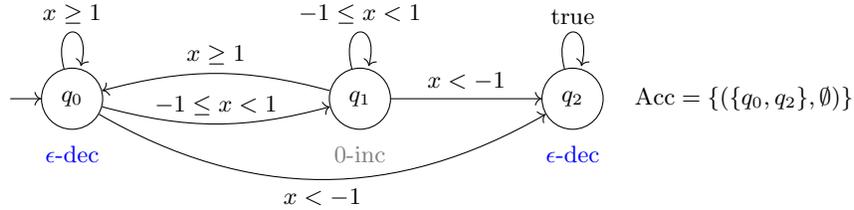
\begin{figure}
\centering
\begin{tikzpicture}[initial text={}]
    \def\edgeshift{1.2cm}
    \node[state, initial] (q0) {$q_0$};
    \node[state,right=3cm of q0] (q1) {$q_1$};
    %\node[state,below=7mm of q1] (q2) {$q_2$};
    \node[state,right=2cm of q1] (q3) {$q_2$};
    \path (q0) edge[bend right=15,->] node[above] {$-1 \leq x < 1$} (q1)
    (q1) edge[bend right=15,->] node[above] {$x \geq 1$} (q0)
    (q1) edge[->] node[above] {$x < -1$} (q3)
    (q1) edge[loop above,->] node[above] {$-1 \leq x < 1$} (q1)
    (q0) edge[loop above,->] node[above] {$x \geq 1$} (q0)
    (q0) edge[bend right=30,->] node[below] {$x < -1$} (q3)
    (q3) edge[loop above,->] node[above] {true} (q3);
    \node[right=3mm of q3] {$\mbox{Acc} = \{ (\{q_0, q_2\}, \emptyset)\}$};

    \node[below=0.1cm of q0] {{\color{blue} $\epsilon$-\text{dec}}};
    \node [below=0.1cm of q1] {{\color{gray} $0$-\text{inc}}};
    \node [below=0.1cm of q3] {{\color{blue} $\epsilon$-\text{dec}}};
\end{tikzpicture}
\caption{Deterministic Streett Automaton for $(x \geq -1) {\sf UG} (-1 \leq x \leq 1)$}
\label{fig:dsa}
\end{figure}
\begin{example}\label{example:UG}
Consider a simple Markov process over one real-valued variable $x$ and control parameter $\kappa$, described by the following stochastic difference equation:
    \begin{equation}
        x_{t+1} = \kappa \cdot x_t + w_t, \qquad x_0 = 100,\qquad w_t \sim {\rm Uniform}(-0.1,0.1)
    \end{equation}
We wish to synthesise a control parameter $\kappa$ for which the process satisfies the {\em stabilize-while-avoid}
property $\Phi = (x \geq -1) {\sf UG} (-1 \leq x \leq 1)$, which requires the system to avoid $x < -1$ until it stabilizes within $(-1 \leq x \leq 1)$. This corresponds to the DSA in \Cref{fig:dsa}, whose states  
define the necessary drift conditions. 

Applying \Cref{thm:requirements-corollary} to the DSA of \Cref{fig:dsa}, we make two observations. Firstly, recalling the intuition of the first trajectory in \Cref{fig:intuition}, we note that the specification is satisfied only if $q_0$ is visited finitely many times by the product process, and this must be established by a Streett supermartingale that strictly decreases when $x \geq 1$ and does not increase otherwise. Secondly, recalling the intuition of the third trajectory, we note that such a Streett supermartingale exists only if $q_2$ is never reached, and this must be established by a supporting invariant. A control parameter for which $\Phi$ is satisfied is $\kappa = 0.5$, and this is established by the following Streett supermartingale and supporting invariant:
\begin{equation}
    V(x,q) = \begin{cases}
        x+1& \text{if }q = q_0\\
        0& \text{otherwise}
    \end{cases}\qquad
    I(x,q) = \begin{cases}
        x \geq -0.2 & \text{if }q = q_0\\
        -0.2 \leq x \leq 0.9 & \text{if } q= q_1\\
        \text{false}& \text{if }q = q_2
    \end{cases}\label{eq:exampleVandI}
\end{equation}
Here, the post-expectation of $V$ results in the function below; note that no term for the stochastic disturbance appears because, in this case, the expected value of $w$ is 0, and so is its contribution to the post-expectation of $V$: 
\begin{equation}
    \Post V(x,q) = \begin{cases}
        0.5 \cdot x + 1&\text{if }x \geq 1 \text{ and } q \in \{ q_0, q_1 \}\\
        0 &\text{otherwise }
    \end{cases}
\end{equation}
Altogether, we obtain the following (satisfied) 
system of universally quantified sentences.
The initiation condition (cf. \cref{eqn:VerifCondition-init}) results in the following sentence:
\begin{align}
    &100 \geq -0.2 \equiv I(x_0, q_0)
\end{align}
The consecution condition (cf.~\cref{,eqn:VerifCondition-consecution}) expands into the following three implications, each of which corresponds to a case of $I$, 
a feasible transition in the automaton, and a feasible transition of the dynamical model according to the sample space ${\cal W} = [-0.1, 0.1]$ of stochastic disturbances:
\begin{align}
    &\forall x \in \mathbb{R}, w \in {\cal W} \colon \underbrace{x \geq -0.2}_{I(x, q_0)}  \wedge \underbrace{x \geq 1}_{\delta(q_0, \cdot) = q_0} \implies \underbrace{0.5 x + w \geq -0.2}_{I(0.5 x + w, q_0)}\\
    &\forall x \in \mathbb{R}, w \in {\cal W} \colon \underbrace{x \geq -0.2}_{I(x, q_0)} \land \underbrace{-1 \leq x < 1}_{\delta(q_0, \cdot)=q_1} \implies \underbrace{-0.2 \leq 0.5 x + w \leq 0.9}_{I(0.5 x + w, q_1)}\\
    &\forall x \in \mathbb{R}, w \in {\cal W} \colon \underbrace{x \geq -0.2}_{I(x, q_0)} \wedge \underbrace{x < -1}_{\delta(q_0, \cdot) = q_2} \implies \underbrace{\rm false}_{I(0.5 x + w, q_2)}\\
    &\forall x \in \mathbb{R}, w \in {\cal W} \colon \underbrace{-0.2 \leq x \leq 0.9}_{I(x, q_1)}\wedge \underbrace{x \geq 1}_{\delta(q_1, \cdot) = q_0} \implies \underbrace{0.5 x + w \geq -0.2}_{I(0.5 x + w, q_0)}\\
    &\forall x \in \mathbb{R}, w \in {\cal W} \colon \underbrace{-0.2 \leq x \leq 0.9}_{I(x, q_1)} \land \underbrace{-1 \leq x < 1}_{\delta(q_1, \cdot) = q_1} \Rightarrow \underbrace{-0.2 \leq 0.5 x + w \leq 0.9}_{I(0.5 x + w, q_1)}\\
    &\forall x \in \mathbb{R}, w \in {\cal W} \colon \underbrace{-0.2 \leq x \leq 0.9}_{I(x, q_1)} \land \underbrace{x < -1}_{\delta(q_1, \cdot) = q_2} \implies \underbrace{\rm false}_{I(0.5 x + w, q_2)}\\
    &\forall x \in \mathbb{R}, w \in {\cal W} \colon \underbrace{\rm false}_{I(x, q_2)} \land \underbrace{\rm true}_{\delta(q_2, \cdot) = q_2} \implies \underbrace{\rm false}_{I(0.5 x + w, q_2)}
\end{align}
The strict decrease drift condition associated with $q_0$ (cf. \cref{eqn:VerifCondition-epsDec}) results, with $\epsilon = 0.5$, in the following two sentences associated with each case of $\Post V$:
\begin{align}
    &\forall x \in \Real \colon \underbrace{x \geq -0.2}_{I(x, q_0)} \wedge (x \geq 1) \implies \underbrace{0.5 \cdot x + 1}_{\Post V(x, q_0)} ~\leq~ \underbrace{x + 1}_{V(x, q_0)} - \underbrace{0.5}_{\epsilon}\\
    &\forall x \in \Real \colon \underbrace{x \geq -0.2}_{I(x, q_0)} \wedge (x < 1) \implies \underbrace{0}_{\Post V(x, q_0)} ~\leq~ \underbrace{x + 1}_{V(x, q_0)} - \underbrace{0.5}_{\epsilon}
\end{align}
Similarly, the non-increase drift condition associated with $q_1$ (cf. \cref{eqn:VerifCondition-0inc}) results in the following two implications:
\begin{align}
    &\forall x \in \Real \colon\underbrace{-0.2 \leq x \leq 0.9}_{I(x, q_1)} \wedge (x \geq 1) \implies \underbrace{0.5 \cdot x + 1}_{\Post V(x, q_1)} ~\leq \underbrace{0}_{V(x, q_1)}\label{eqn:impossible-q1}\\
    &\forall x \in \Real \colon \underbrace{-0.2 \leq x \leq 0.9}_{I(x, q_1)} \wedge (x < 1) \implies \underbrace{0}_{\Post V(x, q_1)} \leq \underbrace{0}_{V(x, q_1)}\label{eqn:drift-q1-second}
\end{align}
We note that the invariant in state $q_1$ is sufficiently strong to exclude the possibility of a transition back to $q_0$ from $q_1$ (which is associated with $x \geq 1$), making the premise of the implication in \cref{eqn:impossible-q1} false.
The drift condition of $q_2$ is also trivially satisfied, as the premise of the respective implication is false:
\begin{align}
    \forall x \in \mathbb{R} \colon \underbrace{\rm false}_{I(x, q_2)} \implies \underbrace{0}_{\Post V(x, q_2)} \leq \underbrace{0}_{V(x, q_2)} - \underbrace{0.5}_{\epsilon}
\end{align}
Finally, the non-negativity condition (cf. \cref{eqn:VerifCondition-NonNeg}) is trivially satisfied on $q_1$ and $q_2$ as $V(x, q_1) = V(x, q_2) = 0$. For $q_0$ instead, the condition is the following:
\begin{align}
    &\forall x \in \mathbb{R} \colon \underbrace{x \geq -0.2}_{I(x, q_0)} \implies \underbrace{x + 1}_{V(x, q_0)} \geq 0  
\end{align}
Notably, every sentence consists of a conjunction of inequalities implying an inequality. As we show in \Cref{sec:synth}, difference equations, Streett supermartingales and supporting invariants that are piecewise-defined according to a template as in \cref{eq:exampleVandI} always result in systems of constraints in this form. This enables effective algorithmic synthesis of Streett supermartingales, supporting invariants and control parameters using symbolic or numerical decision procedures.
\qed
\end{example}

\section{Algorithmic Synthesis of Streett Supermartingales}\label{sec:synth}

Exhibiting Streett supermartingales and supporting invariants constitutes a witness that the stochastic dynamical model and its control parameter comply with the $\omega$-regular property at hand. Under the assumption that these three objects are constrained to be in the form of a template, the verification and control problem is reducible to a decision procedure for quantified first-order formulae.
In this section, we define templates that allow effective synthesis using standard symbolic and numerical decision procedures. 

We show that under different assumptions and problem settings, the verification and control problem reduces to the following decision procedures:
\begin{description}
    \item[General Control] This refers to the general synthesis of a Streett supermartingale, supporting invariant, and control parameters. When these and the associated post-expectation are in piecewise polynomial form (\Cref{ssec:pw-poly-systems}), then the synthesis problem is reducible to a quantified formula (with one quantifier alternation) in non-linear real arithmetic (NRA). 
    When they are in piecewise linear form (\Cref{sec:pwl-parametric}) then the synthesis problem reduces to the existential theory of non-linear real arithmetic (QF\_NRA). 

    \item[Shielded Control] This refers to the synthesis of a Streett supermartingale and control parameter, given an externally provided inductive invariant. Externally provided invariants are relevant when a shield that ensures the safety of the policy (but not necessarily its liveness) is computed beforehand~\cite{DBLP:conf/aaai/AlshiekhBEKNT18}.
    This reduces to quadratically constrained programming (QCP) with piecewise linear templates (\Cref{ssec:pwl-constant-guard,example:shielded-control}).
    
    \item[Verification] This refers to the sole synthesis of Streett supermartingales, when the system has a known invariant that is provided a priori. This reduces to linear programming (LP) when templates and post-expectation are piecewise linear (\Cref{ssec:pwl-constant-guard,example:verification}).
\end{description}

We introduce a functional template $F \colon \Real^{N} \times Q \times \Lambda \to \Real$ that maps an $N$-dimensional real-valued vector, 
a state of the automaton $q \in Q$, and a generic template parameter $\lambda \in \Lambda$ to a real-valued output according to a number of cases, guarded by logical predicates:
\begin{equation}\label{eqn:template-T}
\begin{aligned}
F(x,q; \lambda) &= \begin{cases}
        g_{1, l+1}(x; \lambda) & \text{if } \bigwedge_{i = 1}^{l} g_{1,i}(x; \lambda) \lesssim_{1,i} 0, \text{ and } q \in Q_1\\
          &\vdots \\
        g_{m, l+1}(x; \lambda) & \text{if } \bigwedge_{i = 1}^{l} g_{m,i}(x) \lesssim_{m,i} 0, \text{ and } q \in Q_m,\\
    \end{cases}
\end{aligned}
\end{equation}
The value $N$ is a placeholder for either the dimensionality of the state space $\Real^n$, or the joint dimensionality of the system and the stochastic disturbance inputs $\Real^n \times {\cal W}$, according to context. The sets $Q_1, \ldots, Q_m \subseteq Q$ denote constraints on 
the automaton states and $\lesssim$ denotes either a strict- or non-strict inequality. This makes the form of \cref{eqn:template-T} suitable as a template for expressing Streett supermartingales $V(x, q; \theta) \equiv F(x, q; \theta)$, supporting invariants $I(x, q ; \eta)  \equiv \bigwedge [F(x, q; \eta) \leq 0]$, 
dynamical models $f(x,w; \kappa) \equiv F((x, w), -; \kappa)$ as well as the symbolic post-expectation $\Post V(x, q ; \theta, \kappa) \equiv F(x, q; \theta, \kappa)$.

Assuming, without loss of generality, that each observable proposition in $\Pi$ (cf.\ \cref{sec:verification}) corresponds to a single inequality over the state space $\Real^n$, the transition function $\delta(q, \llangle x \rrangle)$ of the automaton takes the form of template $D : \Real^n \times Q \to Q$: 
\begin{equation}\label{eqn:template-D}
\begin{aligned}
    D(x, q) &= 
    \begin{cases}
        q_1' & \text{if } \bigwedge_{i = 1}^{l} g_{1,i}(x) \lesssim_{1,i} 0, \text{ and } q = q_1\\
          &\vdots \\
        q_m' & \text{if } \bigwedge_{i = 1}^{l} g_{m,i}(x) \lesssim_{m,i} 0, \text{ and } q = q_m.\\
    \end{cases}
\end{aligned}
\end{equation}
where each of the automaton's transitions corresponds to a case of \cref{eqn:template-D}. 

The requirements of \cref{eqn:VerifCondition-init,eqn:VerifCondition-consecution,eqn:VerifCondition-epsDec,eqn:VerifCondition-Minc,eqn:VerifCondition-0inc,eqn:VerifCondition-NonNeg} reduce to a conjunction of sentences of the form \cref{eqn:primal-implications-form-only-one}, namely, a universally quantified implication over $N$-dimensional real-valued variables, where each implication has a premise that is a finite conjunction of inequalities (where $L$ is a placeholder for the number of conjuncts), and a consequent that is a single non-strict inequality:
% {\color{red} what is m? Does not match with usage above. Also, what is L?}
\begin{equation}\label{eqn:primal-implications-form-only-one}
\begin{aligned}
    \forall y \in \Real^{N} \colon
    \bigwedge_{i = 1}^L g_{i}(y; \lambda) \lesssim_{i} 0 &\implies g_{L+1}(y; \lambda) \leq 0, 
\end{aligned}
\end{equation}
This is because our construction only invokes compositions of the templates $F$ and $D$ that produce results that are representable in the form of template $F$, namely, a piecewise function over $\Real^N \times Q$ with parameters $\lambda \in \Lambda$. 
In combination with rewriting at the level of propositional logic, we establish \cref{eqn:primal-implications-form-only-one}.

Finally, we note that the conjunction of sentences of the form \cref{eqn:primal-implications-form-only-one} is existentially quantified over the certificate, invariant and control parameters, as well as the parameters $\epsilon$ and $M$, all of which we notationally subsume within $\lambda$.
We now discuss algorithms for finding a satisfying assignment to these existentially quantified parameters under the problem scenarios outlined earlier.

\subsection{Piecewise Polynomial Systems and Templates}\label{ssec:pw-poly-systems}

Under the assumption that all functions $g$ in the templates \cref{eqn:template-T,eqn:template-D} are polynomials in \( x \in \Real^N \) and $ \lambda \in \Lambda $, the synthesis problem reduces to an existentially quantified conjunction of statements in the form of \cref{eqn:primal-implications-form-only-one}, which are in turn universally quantified implications over polynomial inequalities. This synthesis problem belongs to the first-order theory of nonlinear real arithmetic (NRA) and is decidable using quantifier elimination~\cite{DBLP:journals/cca/Collins74}.

\subsection{Piecewise Linear Systems \& Templates with Parametric Guards}\label{sec:pwl-parametric}

Despite its decidability, the decision procedures for NRA are computationally feasible only for small problems. By making additional assumptions about the system dynamics and templates, we improve the feasibility of the synthesis problem using Farkas' Lemma.
The Farkas' Lemma \cite[p.32 \& Table 2.4.1, p.34]{NonlinearMangasarian} states that the following two sentences are equivalent:
\begin{align}
\label{eqn:primal-matrix-form}
    &\forall y \in \Real^{N} \colon  Ay \leq b \implies c^{\sf T}y \leq d\\
    &\exists z \in \Real^L_{\geq 0} \colon 
    \left(\begin{array}{l}
A^{\sf T} z = c\\
\wedge ~b^{\sf T}z \leq d 
\end{array}\right) \vee \left(\begin{array}{l}
A^{\sf T}z = 0\\
\wedge~ b^{\sf T}z < 0
\end{array}\right)\label{eqn:after-farkas-general}
\end{align}
with $z$ constituting a freshly introduced set of variables.  
This rewrite eliminates the quantifier alternation and yields a decision problem in the first-order existential theory of non-linear real arithmetic (QF\_NRA). In the case where the functions $g$ in \cref{eqn:primal-implications-form-only-one} are linear in the variables $y \in \Real^{N}$, 
and with the help of a technical result that allows strict inequalities in \cref{eqn:primal-implications-form-only-one} to be replaced by non-strict inequalities (cf.\ \cite[Lemma 1]{DBLP:journals/toplas/ChatterjeeFNH18}), we find that
\cref{eqn:primal-implications-form-only-one} takes the form of \cref{eqn:primal-matrix-form}, allowing Farkas' Lemma to be applied.

\begin{example}[General Control]\label{example:general-control}
Considering \cref{example:UG}, suppose we want to synthesise a value for the control parameter $\kappa$ such that the specification $\Phi$ is satisfied almost surely, along with a Streett supermartingale and supporting invariant. For this purpose, we introduce template parameters $\theta = (\alpha_0, \beta_0, \alpha_1, \beta_1, \alpha_2, \beta_2)$, $\eta = (\eta_1, \eta_2, \eta_3, \eta_4)$ and template the Streett supermartingale and supporting invariant using the following form:
\begin{equation}
\begin{aligned}
    V(x,q_1; \theta) &= 
        \alpha_1 \cdot  x+ \beta_1 \\
    I(x,q_1; \eta) &= 
        (\eta_1 \cdot x  \leq \eta_2) \land 
        (\eta_3 \cdot x  \leq \eta_4)
    \label{eq:exampleVandITemplates}
\end{aligned}
\end{equation}
proceeding analogously for states other than $q_1$,
which yields for $q \in \{q_0, q_1, q_2\}$ the following expression for $\Post V$ in terms of the control parameter $\kappa$:
\begin{equation}
\begin{aligned}
    \Post V(x, q; \theta, \kappa) &= 
    \begin{cases}
        \alpha_0 \kappa \cdot x + \beta_0 &\text{if }x \geq 1\\
        \alpha_1 \kappa \cdot x + \beta_1 &\text{if }-1 \leq x < 1\\
        \alpha_2 \kappa \cdot x + \beta_2 &\text{if } x < -1
    \end{cases}
\end{aligned}
\end{equation}
Substituting these expressions into \cref{eqn:VerifCondition-init,eqn:VerifCondition-consecution,eqn:VerifCondition-epsDec,eqn:VerifCondition-Minc,eqn:VerifCondition-0inc,eqn:VerifCondition-NonNeg} results in a conjunction of implications of the form \cref{eqn:primal-implications-form-only-one} over inequalities that are linear in the variable $x \in \Real$, but polynomial over the existentially quantified parameters. For example, the non-increasing drift condition associated with $q_1$ (cf. \cref{eqn:VerifCondition-0inc,eqn:impossible-q1,eqn:drift-q1-second}) corresponds to a number of implications, one for each case of the piecewise-defined $\Post V(x, q_1; \theta, \kappa)$. Considering the case $x \geq 1$, we see that the templated implication analogous to \cref{eqn:impossible-q1} is:
\begin{equation}\label{eqn:impossible-q1-templated}
\begin{aligned}
    &\forall x \in \Real \colon
    \underbrace{
    \begin{bmatrix}
\eta_1  \\
\eta_3 \\
-1
\end{bmatrix}
    \begin{bmatrix}
x
\end{bmatrix}
\leq 
    \begin{bmatrix}
    \eta_2  \\
    \eta_4  \\
-1
\end{bmatrix}
}_{I(x, q_1; \eta) \wedge (x \geq 1)}
    \implies \underbrace{
    \begin{bmatrix}
    \alpha_0 \kappa - \alpha_1
    \end{bmatrix}
    \begin{bmatrix}
    x
    \end{bmatrix}
    \leq    
\begin{bmatrix}
    \beta_1 - \beta_0
    \end{bmatrix}
    }_{
    \Post V(x, q_1; \theta, \kappa) \leq V(x,q_1; \theta)}
\end{aligned}
\end{equation}
which is in the form of \cref{eqn:primal-matrix-form} and yields an existentially quantified disjunction of polynomial inequalities (over the existentially quantified variables, which include the template and control parameters) once rewritten into form \cref{eqn:after-farkas-general}, namely a problem in the existential first-order theory of non-linear real arithmetic. \qed
\end{example}

\subsection{Piecewise Linear Systems and Templates with Known Guards}\label{ssec:pwl-constant-guard}

Supposing additionally that an inductive invariant is externally provided, we further improve the computational feasibility of the synthesis problem by reducing it to a quadratically-constrained programming (QCP) problem.
In this setting, all inequalities in the premise of \cref{eqn:primal-implications-form-only-one} are known linear inequalities of the vector $y$, and the matrix $A$ and vector $b$ in \cref{eqn:primal-matrix-form} are constant (i.e.\ contain no existentially quantified variables). Therefore, the satisfiability of the premise of \cref{eqn:primal-matrix-form} is decidable using linear programming to check whether $Ay \leq b$ admits any solution for $y$. After removing any implications of the form \cref{eqn:primal-matrix-form} which possess an unsatisfiable premise, we may exploit a special case of Farkas' Lemma that assumes a satisfiable premise \cite[Theorem 3]{DBLP:journals/toplas/ChatterjeeFNH18}. This version states that if there exists a solution to the system $Ay \leq b$, then the formula \cref{eqn:primal-matrix-form} is equivalent to
\begin{equation}\label{eqn:farkas-premise-sat-dual}
\exists z \in \Real^L_{\geq 0} \colon 
    \left(\begin{array}{l}
A^{\sf T} z = c\\
\wedge ~b^{\sf T}z \leq d
\end{array}\right).
\end{equation}
This formula is an existentially quantified conjunction of inequalities, thus transforming the synthesis problem into deciding the satisfiability of a conjunction of polynomial constraints. Such a system of polynomial constraints is reducible to QCP, since higher degree polynomial expressions may be constructed from quadratic constraints by introducing fresh variables. This establishes the reduction to QCP for \textit{shielded control} when applied to piecewise linear systems and templates, with known invariant. Furthermore, as illustrated in \cref{example:verification}, if additionally the system is autonomous, the synthesis problem reduces to an LP.

\begin{example}[Shielded Control]\label{example:shielded-control}
Continuing from \cref{example:general-control}, we note that if a sufficiently strong invariant is provided a priori (such as that of \cref{eq:exampleVandI}), then the synthesis problem reduces to implications of the form \cref{eqn:primal-implications-form-only-one} with the property that the linear inequalities occurring within the premise of an implication have constant coefficients. Instead of \cref{eqn:impossible-q1-templated}, for example, we obtain:
\begin{equation}\label{eqn:impossible-q1-templated-known-inv}
\begin{aligned}
    &\forall x \in \Real \colon
    \underbrace{
    \begin{bmatrix}
1  \\
-1 \\
-1
\end{bmatrix}
    \begin{bmatrix}
x
\end{bmatrix}
\leq 
    \begin{bmatrix}
0.9  \\
0.2  \\
-1
\end{bmatrix}
}_{I(x, q_1; 1, 0.9, -1, 0.2 ) \wedge (x \geq 1) }
%\text{\eqref{eq:exampleVandI}
    \implies \underbrace{
    \begin{bmatrix}
    \alpha_0 \kappa - \alpha_1
    \end{bmatrix}
    \begin{bmatrix}
    x
    \end{bmatrix}
    \leq    
\begin{bmatrix}
    \beta_1 - \beta_0
    \end{bmatrix}
    }_{
    \Post V(x, q_1; \theta, \kappa) \leq V(x,q_1; \theta)}
\end{aligned}
\end{equation}
The premise of \cref{eqn:impossible-q1-templated-known-inv} is a known system of linear inequalities, so if its premise is satisfiable (decidable via linear programming) an application of \cref{eqn:farkas-premise-sat-dual} transforms the synthesis problem into an existentially quantified conjunction of polynomial constraints. The particular constraint \cref{eqn:impossible-q1-templated-known-inv} has an unsatisfiable premise, however, and is thus vacuously true. \qed
\end{example}

\begin{example}[Verification]\label{example:verification}
Assuming that $\kappa = 0.5$, the dynamical model results in an autonomous system, and if a sufficiently strong supporting invariant is provided a priori (as is precisely the case in \cref{example:UG}), then the implication \cref{eqn:impossible-q1-templated-known-inv} becomes:
\begin{equation}\label{eqn:impossible-q1-templated-Vonly}
\begin{aligned}
    &\forall x \in \Real \colon
    \underbrace{
    \overbrace{
    \begin{bmatrix}
1  \\
-1 \\
-1
\end{bmatrix}
}^{A}
    \begin{bmatrix}
x
\end{bmatrix}
\leq \overbrace{
    \begin{bmatrix}
0.9  \\
0.2  \\
-1
\end{bmatrix}}^{b}
}_{I(x, q_1; 1, 0.9, -1, 0.2) \wedge (x \geq 1)}
    \implies \underbrace{
    \overbrace{\begin{bmatrix}
    0.5 \cdot \alpha_0 - \alpha_1
    \end{bmatrix}}^{c^{\sf T}}
    \begin{bmatrix}
    x
    \end{bmatrix}
    \leq    
    \overbrace{
    \begin{bmatrix}
    \beta_1 - \beta_0
    \end{bmatrix}
    }^{d}
    }_{
    \Post V(x, q_1; \theta, 0.5) \leq V(x,q_1; \theta)}
\end{aligned}
\end{equation}
In this case, for an implication with satisfiable premise, we may apply \cref{eqn:farkas-premise-sat-dual} to obtain an existentially quantified conjunction of inequalities that are linear in $x$, but further note that matrix $A$ and vector $b$ have constant entries, whereas the vector $c$ and scalar $d$ are linear expressions over template variables. Thus, an application of \cref{eqn:farkas-premise-sat-dual} generates an existentially quantified conjunction of linear constraints, which is decidable using a linear program. \qed
\end{example}

\section{Experimental Evaluation}\label{sec:experiments}

We implement our algorithmic technique for the synthesis of Streett supermartingales, supporting invariants, and control policies. Our implementation does not require externally provided invariants, and assumes a template for the Streett supermartingale that is linear in the state variables, and thereby of the form \cref{eqn:template-T} with a single case for each automaton state. We assume a convex polyhedral template for the supporting invariant, and apply Farkas' Lemma to produce a decision problem in QF\_NRA (\Cref{sec:pwl-parametric}). In \Cref{tab:experiments}, we demonstrate examples of $\omega$-regular properties and of infinite-state probabilistic systems, with piecewise linear dynamics, certificates and supporting invariants. The \textit{Output} column of \Cref{tab:experiments} describes the synthesis problem (cf.\ \cref{sec:synth}): VIC indicates \textit{general control}; VC indicates \textit{shielded control}; VI indicates synthesis of Streett supermartingales and supporting invariants for an autonomous stochastic dynamical model; V indicates \textit{verification} (namely, synthesis of only Streett supermartingales, using an externally provided invariant).

We use the Spot library \cite{DBLP:conf/cav/Duret-LutzRCRAS22} to translate the LTL formulae shown in \Cref{tab:experiments} into deterministic Streett automata with state-based acceptance conditions. We use SymPy \cite{DBLP:journals/peerjpre/MeurerSPCRK0MSR16} to perform symbolic manipulations and generate the corresponding decision problem, which we solve using an off-the-shelf SMT solver, Z3 \cite{DBLP:conf/tacas/MouraB08,DBLP:journals/cca/JovanovicM12}. The systems in \Cref{tab:experiments} are all infinite-state, namely continuous-state models, with the exception of \texttt{evenOrNegative} that has a countably-infinite state space. The benchmarks make use of discrete random variables, which allows for the post-expectation to be calculated by weighted enumeration of probabilistic choices in the product process (with the exception of {\tt evenOrNegative}, for which the post-expectation is provided manually). Since our implementation entails deterministic algorithms, we provide the time associated with a single execution owing to negligible variance in these timings.

\begin{table}[]
    \centering
    \caption{Output of our experiments for a range of infinite-state probabilistic systems and $\omega$-regular properties.} 
    \begin{tabular}{l||c|l|r}
    Benchmark & $\omega$-Regular Specification & Output & Time [s] \\
\hline

    {\tt evenOrNegative} (Fig.~\ref{fig:simple}) & ${\sf GF}(x \text{ even}) \vee {\sf FG}(x < 0)$ & V & 0.09 \\  

    {\tt SafeRWalk1} & ${\sf G}(x < 100)$ & VIC & 1.09 \\

    {\tt PersistRW} & ${\sf FG}(x \leq 10)$ & VI & 1.16 \\
    
    {\tt RecurRW} & ${\sf GF}(x > 100)$ & VI & 1.49 \\

    {\tt SafeRWalk2} & ${\sf G}(x \geq 10)$ & VIC & 1.09 \\

    {\tt GuaranteeRW} & ${\sf G}(x \geq -10) \rightarrow {\sf F}(x \geq 10^3)$ & VI & 5.61 \\

    {\tt Temperature1} & {\sf FG}($\neg \text{Hot} \wedge \neg \text{Cold}$) & VIC & 4.11 \\

    {\tt Temperature2} & ${\sf GF}(x \leq 30) \wedge {\sf G}(x \leq 60)$ & VI & 28.93 \\

    {\tt Temperature3} & ${\sf G}(\text{Safe}) \wedge 
    \left[  {\sf GF}(\text{Cold}) \rightarrow 
    {\sf GF}(\text{Hot})\right]$ & VIC & 28.58 \\

    {\tt Temperature4} & ${\sf G}(\text{Safe}) \wedge 
    \left[  {\sf GF}(\text{Cold}) \rightarrow 
    {\sf GF}(\text{Hot})\right]$ & VC & 4.64 \\

    {\tt FinMemoryControl} & ${\sf GF}(x \leq 0) \rightarrow {\sf GF}(x \geq 100)$ & VIC & 16.73 \\
    \end{tabular}
    
    \label{tab:experiments}
\end{table}

The benchmark {\tt Temperature1} (\Cref{tab:experiments}) is an instance of a \textit{general control} problem (\Cref{sec:synth}) that models an air-conditioned room that dissipates heat to its surroundings (at temperature $x_\text{ext}$), with stochastic fluctuations of the room temperature. The state $x_t \in \Real$ is the temperature of the room, $x_\text{ext} = 280\text{K}$ is the external temperature, and the desired temperature is $x_\text{set}=295\text{K}$, with $x_0 = x_\text{ext}$. The system has the following dynamics:
\begin{equation}\label{eqn:temp-model-dynamics}
    x_{t+1} = x_{t} - \frac{1}{100} (x_{t} - x_\text{ext}) + (\alpha x_t + \beta) + \frac{1}{10}(2 w_t - 1), 
\end{equation}
with $w_t \sim \text{Bernoulli}(0.5)$. The dynamics depend upon the parameters $\alpha, \beta$ of the controller, restricted to $\alpha, \beta \in [-10, 10]$, to reflect the capabilities of the controller.
We define two observations, $\Pi = \{ \text{Hot}, \text{Cold} \}$ with $\llangle x \rrangle = \{ \text{Hot} \}$ when the temperature $x$ exceeds $x_\text{set} + 3$, $\llangle x \rrangle = \{ \text{Cold} \}$ when the temperature falls below $x_\text{set} - 3$, and $\llangle x \rrangle = \emptyset$ otherwise. The specification is ${\sf FG}(\neg \text{Hot} \wedge \neg \text{Cold})$, namely, that the temperature eventually persists within the interval $(292, 298)$ around $x_\text{set} = 295$. Our method synthesises a certificate, supporting invariant, and controller with $\alpha = -1/32, \beta = 4787/512$. The supporting invariant is a conjunction of two linear inequalities at each automaton state.

We next illustrate how shielding improves the efficiency of our synthesis algorithm. {\tt Temperature3} involves the same controlled dynamics as \cref{eqn:temp-model-dynamics}, but we add a new observation $\{ \text{Safe} \}$ associated with the temperature being under 310K, and aim to satisfy the property ${\sf G}(\text{Safe}) \wedge ({\sf GF}(\text{Cold}) \rightarrow {\sf GF}(\text{Hot}))$. To synthesise a memoryless controller $\alpha = -1/64, \beta = 9/2$ for {\tt Temperature3} along with a suitable inductive invariant requires a total of 28.58s (of which the QF\_NRA solver requires 23.65s). In {\tt Temperature4} we consider a shielded memoryless controller that ensures the temperature always stays under 310K:
\begin{equation}\label{eqn:shielded-temperature-dynamics}
\begin{aligned}
    x_{t+1} &= 
    x_{t} - \frac{1}{100} (x_{t} - x_\text{ext})
    + \frac{1}{10}(2 w_t - 1)
    +
    \begin{cases}
    \alpha x_{t} + \beta &x_{t} < 305\\
    -3 &x_{t} \geq 305
    \end{cases}
\end{aligned}
\end{equation}
and we desire a certified controller for the same reactivity property as in benchmark {\tt Temperature3}.
We constrain $\alpha,\beta \in [-5, 5]$ (as modelling assumptions), and we impose $305 \cdot \alpha + \beta < 5.24$ to ensure that the temperature never exceeds 310K. We provide an invariant a priori, and the resulting QCP is solvable in 0.03s to obtain $\alpha=-43/3200, \beta=4$.

To illustrate how our framework is applicable to finite memory controllers, we consider in {\tt FinMemoryControl} (\Cref{tab:experiments}) a controller that has one bit of memory (denoted by $b \in \{0, 1\}$), which is updated according to the current state $x$. That is, the system has state space $\Real \times \{ 0, 1 \}$, with update function (cf.\ \cref{eqn:update-func}):
\begin{equation}\label{eqn:finmemorytemplate}
    f(x,b,w; \kappa) = 
    \begin{cases}
        (x + \alpha + w, 1) &\text{if }b = 1 \wedge l\cdot x \leq m\\
        (x + \alpha + w, 0)&\text{if }b = 1 \wedge l \cdot x > m\\
        (x + w - 1, 1)&\text{if }b = 0 \wedge p \cdot x \leq q \\
        (x + w - 1, 0) &\text{if }b = 0 \wedge p \cdot x > q
    \end{cases}
\end{equation}
where $\kappa =  ( l, m, p, q, \alpha )$ is the set of control parameters. That is, we wish to synthesise the output of the controller, $\alpha$, but also the logic that determines how the controller's memory is to be updated, given the reactivity specification ${\sf GF}(x \leq 0) \rightarrow {\sf GF}(x \geq 100)$. This is a decision problem in QF\_NRA given that the guards of the template for $f(x, b, w; \kappa)$ contain template parameters (\Cref{sec:pwl-parametric}). Our method finds $\alpha=56, l=1/8,m=14,p=1/2,q=51$.

In summary, we find that our synthesis procedure based on Farkas' Lemma (\Cref{sec:pwl-parametric}) allows for the practical synthesis of Streett supermartingales, supporting invariants and control parameters for a range of $\omega$-regular properties for infinite (countable/continuous) state piecewise linear probabilistic systems, with our tool terminating in under 30s for the examples considered. We further illustrate that stronger assumptions (e.g.\ the external provision of shields or supporting invariants) improve the computational efficiency of control synthesis (cf.\ {\tt Temperature3} vs.\ {\tt Temperature4}).

\section{Related Work}\label{sec:related}

 The verification of finite Markov chains is a classic topic for which automated and scalable algorithmic tools exist \cite{DBLP:conf/cav/KwiatkowskaNP11}, which combine graph algorithms and linear algebra to directly compute the probability of satisfying an $\omega$-regular specification \cite{DBLP:books/daglib/0020348}. This approach exploits the limit behaviour of finite Markov chains, reducing the problem to computing the reachability probabilities of bottom strongly connected components by leveraging the finite graph structure. These techniques do not, however, apply to probabilistic processes over countably infinite or continuous state spaces, which are the focus of this work.

Verification of continuous-state Markov processes has been addressed via two main strategies \cite{LSAZ22}. The first approximates the continuous-state process with an abstract finite state process (e.g., through state space discretisation) and performs finite-state model checking on the abstraction \cite{DBLP:conf/qest/DesharnaisLT08,DBLP:journals/ejcon/0001SRHH12,SA13,DBLP:conf/hybrid/TkachevA13}. The second strategy certifies the property of interest by providing a suitable certificate, using supermartingale theory to analyse Markov chains over general state spaces \cite[Chapter 8.4]{meyn_tweedie_glynn_2009}. This includes supermartingale certificates for specific linear-time properties including almost-sure reachability \cite{DBLP:conf/cav/ChakarovS13}, probabilistic safety \cite{DBLP:conf/cav/ChatterjeeGMZ22,DBLP:conf/popl/ChatterjeeNZ17,DBLP:journals/toplas/TakisakaOUH21}, reach-avoidance \cite{DBLP:conf/tacas/ChatterjeeHLZ23}, persistence, and recurrence \cite{DBLP:conf/aaai/LechnerZCH22,DBLP:conf/tacas/ChakarovVS16}. These rules are justified using martingale theory, including concentration inequalities (e.g.\ Azuma's inequality \cite{DBLP:conf/popl/ChatterjeeNZ17}) and the supermartingale convergence theorem \cite[Theorem 5.2.9, p.236]{Durrett2010}, with recent order-theoretic justifications \cite{DBLP:journals/toplas/TakisakaOUH21}.

Although prior work introduced supermartingale proof rules for almost-sure persistence and recurrence \cite{DBLP:conf/tacas/ChakarovVS16}, these are too conservative for general reactivity properties. For instance, in \Cref{fig:simple}, a reactivity property (a disjunction of persistence and recurrence requirements) holds almost surely, even though neither disjunct holds almost surely. Recent work \cite{anand2024compositional} addressed proving $\omega$-regular properties with deterministic Streett automata by synthesising a control policy and barrier certificates for the persistence component of each Streett pair. However, as the authors mention, this approach disregards the recurrence component and is thus conservative and mainly suited to safety specifications \cite[Section 8.1]{anand2024compositional}.

Our approach, by contrast, applies to general Streett pairs, using the Robbins~\&~Siegmund convergence theorem (\Cref{thm:robbins-siegmund}) to establish that the disjunction of persistence and recurrence properties in each Streett pair is satisfied with probability one, without requiring either disjunct to hold almost surely. While the Robbins \& Siegmund convergence theorem has applications in statistics \cite{Anbar1976}, stochastic optimisation \cite[Theorem 17.15]{DBLP:books/daglib/0034861}, and reinforcement learning \cite[Proposition 4.2]{books/lib/BertsekasT96}, we are the first to apply it to derive a supermartingale certificate for Streett conditions and, as a result, $\omega$-regular properties.

The algorithmic synthesis of supermartingale certificates for reachability, probabilistic safety, persistence, and recurrence has been addressed for affine programs and certificates using Farkas' Lemma \cite{AgrawalC018,DBLP:conf/cav/ChakarovS13,DBLP:journals/toplas/ChatterjeeFNH18,DBLP:conf/popl/ChatterjeeNZ17} and for polynomial programs and certificates using Putinar's Positivstellensatz \cite{DBLP:conf/cav/ChatterjeeGMZ22,DBLP:conf/cav/ChatterjeeFG16}, producing linear- or quadratically-constrained programs, assuming suitably strong inductive invariants are provided a priori. These techniques were first introduced for the synthesis of ranking certificates and invariants for deterministic programs \cite{DBLP:conf/tacas/ColonS01,DBLP:conf/cav/ColonS02}. We apply these techniques to synthesise Streett supermartingales, supporting invariants, and control policies by deciding the satisfiability of a single query in the existential first-order theory of reals (\Cref{sec:experiments}), or by solving a QCP or LP when suitable invariants are externally provided (\Cref{ssec:pwl-constant-guard,ssec:pwl-constant-guard}), which may be derived from a shield associated with the controller \cite{DBLP:conf/aaai/AlshiekhBEKNT18}, or when a polyhedral enclosure for the reachable states is known or computed a priori with other methods~\cite{DBLP:books/daglib/0080029,DBLP:conf/cav/ColonSS03,DBLP:journals/scp/ErnstPGMPTX07,DBLP:journals/tosem/NguyenKWF14,DBLP:conf/sas/SankaranarayananSM04}.
Our approach to the joint synthesis of certificates and supporting invariants is in principle applicable to other supermartingale notions studied by prior work \cite{DBLP:conf/tacas/ChakarovVS16,DBLP:journals/toplas/TakisakaOUH21,AgrawalC018,DBLP:conf/cav/ChatterjeeGMZ22}.

The problem of certified control synthesis in infinite state Markov decision processes (MDPs) has been addressed for specific objectives, such as reachability-reward objectives in countable-state MDPs \cite{DBLP:journals/corr/abs-2311-06889} and reach-avoid specifications for continuous-state MDPs \cite{DBLP:conf/tacas/ChatterjeeHLZ23,DBLP:conf/tacas/WangZ23,DBLP:conf/cav/YangZZTPZ23}, as well as specific infinite-horizon properties \cite{TA14,TMKA17}. Here, we provide an automated synthesis approach (\Cref{sec:synth}) applicable to general reactivity properties over continuous-state stochastic processes.

Further automata-theoretic approaches such as recursive Markov chains (RMC) \cite{DBLP:conf/qest/YannakakisE05,DBLP:journals/jacm/EtessamiY09} and probabilistic pushdown automata (pPDA) \cite{DBLP:journals/lmcs/KuceraEM06,DBLP:conf/fossacs/WinklerGK22,DBLP:journals/fmsd/BrazdilEKK13} provide means for specifying stochastic processes over countably infinite state spaces. The $\omega$-regular model checking problem for these has been studied which, under some restrictions on the model, is decidable \cite{DBLP:conf/stacs/BrazdilKS05}.
By contrast, our Streett supermartingale theorems (\cref{sec:streett-sm}) apply to general stochastic processes (including over uncountably infinite state spaces), though identifying the class of $\omega$-regular properties and stochastic processes for which Streett supermartingales are complete (analogous to the notion of positive almost-sure termination \cite{DBLP:conf/popl/FioritiH15}) remains an open problem. However, the algorithms in \cref{sec:synth} are relatively complete: if a Streett supermartingale in linear or polynomial form with a known degree exists, our algorithm will compute it.

\section{Conclusion}

We have introduced the first supermartingale certificate for $\omega$-regular properties, by exploiting the Robbins \& Siegmund convergence theorem applied to deterministic Streett automata. Our result is the most expressive supermartingale certificate to date, 
enabling effective almost-sure verification of reactivity properties without requiring each persistence and recurrence component 
to hold with probability one, as in previous work. 
We have provided an algorithm to reduce the problem of synthesising Streett supermartingales along with supporting inductive invariants and control policies to symbolic (SMT) and numerical (QCP, LP) decision procedures, and have demonstrated the practical efficacy of our method on several verification and control examples. 

Our approach lends itself to extension towards quantitative verification~\cite{DBLP:conf/cav/ChatterjeeGMZ22,DBLP:journals/toplas/TakisakaOUH21}, and towards effective algorithmic synthesis of supermartingale certificates via Positivstellensatz~\cite{DBLP:conf/cav/ChatterjeeFG16}. Furthermore, it is open to data-driven techniques along the lines of recent work on neural certificate learning~\cite{DBLP:conf/nips/ChangRG19,DBLP:journals/csysl/AbateAGP21,DBLP:conf/cav/AbateGR20,DBLP:conf/sigsoft/GiacobbeKP22,DBLP:conf/concur/AbateEGPR23,DBLP:conf/tacas/ChatterjeeHLZ23,DBLP:conf/aaai/NadaliM0024}.

\section*{Acknowledgments}

This research was supported in part by the EPSRC Doctoral Training Partnership, and the Department of Computer Science Scholarship at the University of Oxford. The conference version of this manuscript appeared in the proceedings of CAV'24.

\bibliographystyle{splncs04}
\bibliography{main}

\begin{thebibliography}{10}
\providecommand{\url}[1]{\texttt{#1}}
\providecommand{\urlprefix}{URL }
\providecommand{\doi}[1]{https://doi.org/#1}

\bibitem{DBLP:journals/csysl/AbateAGP21}
Abate, A., Ahmed, D., Giacobbe, M., Peruffo, A.: Formal synthesis of {Lyapunov Neural Networks}. {IEEE} Control. Syst. Lett.  \textbf{5}(3),  773--778 (2021)

\bibitem{DBLP:conf/concur/AbateEGPR23}
Abate, A., Edwards, A., Giacobbe, M., Punchihewa, H., Roy, D.: Quantitative verification with neural networks. In: {CONCUR}. LIPIcs, vol.~279, pp. 22:1--22:18. Schloss Dagstuhl - Leibniz-Zentrum f{\"{u}}r Informatik (2023)

\bibitem{DBLP:conf/cav/AbateGR20}
Abate, A., Giacobbe, M., Roy, D.: Learning probabilistic termination proofs. In: {CAV} {(2)}. Lecture Notes in Computer Science, vol. 12760, pp. 3--26. Springer (2021)

\bibitem{bisimulationlearning}
Abate, A., Giacobbe, M., Schnitzer, Y.: Bisimulation learning. In: {CAV}. Lecture Notes in Computer Science, Springer (2024)

\bibitem{DBLP:journals/ejcon/AbateKLP10}
Abate, A., Katoen, J., Lygeros, J., Prandini, M.: Approximate model checking of stochastic hybrid systems. Eur. J. Control  \textbf{16}(6),  624--641 (2010)

\bibitem{AgrawalC018}
Agrawal, S., Chatterjee, K., Novotn{\'{y}}, P.: Lexicographic ranking supermartingales: an efficient approach to termination of probabilistic programs. Proc. {ACM} Program. Lang.  \textbf{2}({POPL}),  34:1--34:32 (2018)

\bibitem{DBLP:conf/aaai/AlshiekhBEKNT18}
Alshiekh, M., Bloem, R., Ehlers, R., K{\"{o}}nighofer, B., Niekum, S., Topcu, U.: Safe reinforcement learning via shielding. In: {AAAI}. pp. 2669--2678. {AAAI} Press (2018)

\bibitem{anand2024compositional}
Anand, M., Lavaei, A., Zamani, M.: Compositional synthesis of control barrier certificates for networks of stochastic systems against $\omega$-regular specifications. Nonlinear Analysis: Hybrid Systems  \textbf{51},  101427 (2024)

\bibitem{Anbar1976}
Anbar, D.: {An Application of a Theorem of Robbins and Siegmund}. The Annals of Statistics  \textbf{4}(5),  1018 -- 1021 (1976)

\bibitem{DBLP:books/daglib/0020348}
Baier, C., Katoen, J.: Principles of model checking. {MIT} Press (2008)

\bibitem{DBLP:journals/corr/abs-2311-06889}
Batz, K., Biskup, T.J., Katoen, J., Winkler, T.: Programmatic strategy synthesis: Resolving nondeterminism in probabilistic programs. Proc. {ACM} Program. Lang.  \textbf{8}({POPL}),  2792--2820 (2024)

\bibitem{books/lib/BertsekasT96}
Bertsekas, D.P., Tsitsiklis, J.N.: Neuro-dynamic programming., Optimization and neural computation series, vol.~3. Athena Scientific (1996)

\bibitem{DBLP:journals/fmsd/BrazdilEKK13}
Br{\'{a}}zdil, T., Esparza, J., Kiefer, S., Kucera, A.: Analyzing probabilistic pushdown automata. Formal Methods Syst. Des.  \textbf{43}(2),  124--163 (2013)

\bibitem{DBLP:conf/stacs/BrazdilKS05}
Br{\'{a}}zdil, T., Kucera, A., Strazovsk{\'{y}}, O.: On the decidability of temporal properties of probabilistic pushdown automata. In: {STACS}. Lecture Notes in Computer Science, vol.~3404, pp. 145--157. Springer (2005)

\bibitem{DBLP:conf/cav/ChakarovS13}
Chakarov, A., Sankaranarayanan, S.: Probabilistic program analysis with martingales. In: {CAV}. Lecture Notes in Computer Science, vol.~8044, pp. 511--526. Springer (2013)

\bibitem{DBLP:conf/sas/ChakarovS14}
Chakarov, A., Sankaranarayanan, S.: Expectation invariants for probabilistic program loops as fixed points. In: {SAS}. Lecture Notes in Computer Science, vol.~8723, pp. 85--100. Springer (2014)

\bibitem{DBLP:conf/tacas/ChakarovVS16}
Chakarov, A., Voronin, Y., Sankaranarayanan, S.: Deductive proofs of almost sure persistence and recurrence properties. In: {TACAS}. Lecture Notes in Computer Science, vol.~9636, pp. 260--279. Springer (2016)

\bibitem{DBLP:conf/nips/ChangRG19}
Chang, Y., Roohi, N., Gao, S.: {Neural Lyapunov Control}. In: NeurIPS. pp. 3240--3249 (2019)

\bibitem{DBLP:conf/cav/ChatterjeeFG16}
Chatterjee, K., Fu, H., Goharshady, A.K.: Termination analysis of probabilistic programs through {P}ositivstellensatz. In: {CAV} {(1)}. Lecture Notes in Computer Science, vol.~9779, pp. 3--22. Springer (2016)

\bibitem{DBLP:journals/toplas/ChatterjeeFNH18}
Chatterjee, K., Fu, H., Novotn{\'{y}}, P., Hasheminezhad, R.: Algorithmic analysis of qualitative and quantitative termination problems for affine probabilistic programs. {ACM} Trans. Program. Lang. Syst.  \textbf{40}(2),  7:1--7:45 (2018)

\bibitem{DBLP:conf/cav/ChatterjeeGMZ22}
Chatterjee, K., Goharshady, A.K., Meggendorfer, T., \v{Z}ikeli\'{c}, D.: Sound and complete certificates for quantitative termination analysis of probabilistic programs. In: {CAV} {(1)}. Lecture Notes in Computer Science, vol. 13371, pp. 55--78. Springer (2022)

\bibitem{DBLP:conf/tacas/ChatterjeeHLZ23}
Chatterjee, K., Henzinger, T.A., Lechner, M., \v{Z}ikeli\'{c}, D.: A learner-verifier framework for neural network controllers and certificates of stochastic systems. In: {TACAS} {(1)}. Lecture Notes in Computer Science, vol. 13993, pp. 3--25. Springer (2023)

\bibitem{DBLP:conf/popl/ChatterjeeNZ17}
Chatterjee, K., Novotn{\'{y}}, P., \v{Z}ikeli\'{c}, D.: Stochastic invariants for probabilistic termination. In: {POPL}. pp. 145--160. {ACM} (2017)

\bibitem{DBLP:journals/cca/Collins74}
Collins, G.E.: Quantifier elimination for real closed fields by cylindrical algebraic decomposition-preliminary report. {SIGSAM} Bull.  \textbf{8}(3),  80--90 (1974)

\bibitem{DBLP:conf/cav/ColonSS03}
Col{\'{o}}n, M., Sankaranarayanan, S., Sipma, H.: Linear invariant generation using non-linear constraint solving. In: {CAV}. Lecture Notes in Computer Science, vol.~2725, pp. 420--432. Springer (2003)

\bibitem{DBLP:conf/tacas/ColonS01}
Col{\'{o}}n, M., Sipma, H.: Synthesis of linear ranking functions. In: {TACAS}. Lecture Notes in Computer Science, vol.~2031, pp. 67--81. Springer (2001)

\bibitem{DBLP:conf/cav/ColonS02}
Col{\'{o}}n, M., Sipma, H.: Practical methods for proving program termination. In: {CAV}. Lecture Notes in Computer Science, vol.~2404, pp. 442--454. Springer (2002)

\bibitem{DBLP:conf/popl/CookGPRV07}
Cook, B., Gotsman, A., Podelski, A., Rybalchenko, A., Vardi, M.Y.: Proving that programs eventually do something good. In: {POPL}. pp. 265--276. {ACM} (2007)

\bibitem{DBLP:conf/qest/DesharnaisLT08}
Desharnais, J., Laviolette, F., Tracol, M.: Approximate analysis of probabilistic processes: Logic, simulation and games. In: {QEST}. pp. 264--273. {IEEE} Computer Society (2008)

\bibitem{DBLP:conf/cav/Duret-LutzRCRAS22}
Duret{-}Lutz, A., Renault, E., Colange, M., Renkin, F., Aisse, A.G., Schlehuber{-}Caissier, P., Medioni, T., Martin, A., Dubois, J., Gillard, C., Lauko, H.: From spot 2.0 to spot 2.10: What's new? In: {CAV} {(2)}. Lecture Notes in Computer Science, vol. 13372, pp. 174--187. Springer (2022)

\bibitem{Durrett2010}
Durrett, R.: Probability: Theory and Examples, 4th Edition. Cambridge University Press (2010)

\bibitem{DBLP:journals/scp/ErnstPGMPTX07}
Ernst, M.D., Perkins, J.H., Guo, P.J., McCamant, S., Pacheco, C., Tschantz, M.S., Xiao, C.: The daikon system for dynamic detection of likely invariants. Sci. Comput. Program.  \textbf{69}(1-3),  35--45 (2007)

\bibitem{DBLP:journals/jacm/EtessamiY09}
Etessami, K., Yannakakis, M.: Recursive markov chains, stochastic grammars, and monotone systems of nonlinear equations. J. {ACM}  \textbf{56}(1),  1:1--1:66 (2009)

\bibitem{DBLP:conf/popl/FioritiH15}
Fioriti, L.M.F., Hermanns, H.: Probabilistic termination: Soundness, completeness, and compositionality. In: {POPL}. pp. 489--501. {ACM} (2015)

\bibitem{gehr2016psi}
Gehr, T., Misailovic, S., Vechev, M.T.: {PSI:} exact symbolic inference for probabilistic programs. In: {CAV} {(1)}. Lecture Notes in Computer Science, vol.~9779, pp. 62--83. Springer (2016)

\bibitem{DBLP:conf/sigsoft/GiacobbeKP22}
Giacobbe, M., Kroening, D., Parsert, J.: Neural termination analysis. In: {ESEC/SIGSOFT} {FSE}. pp. 633--645. {ACM} (2022)

\bibitem{DBLP:journals/sttt/HenselJKQV22}
Hensel, C., Junges, S., Katoen, J., Quatmann, T., Volk, M.: The probabilistic model checker {Storm}. Int. J. Softw. Tools Technol. Transf.  \textbf{24}(4),  589--610 (2022)

\bibitem{DBLP:journals/cca/JovanovicM12}
Jovanovic, D., de~Moura, L.: Solving non-linear arithmetic. {ACM} Commun. Comput. Algebra  \textbf{46}(3/4),  104--105 (2012)

\bibitem{DBLP:conf/cav/KretinskyMSZ18}
Kret{\'{\i}}nsk{\'{y}}, J., Meggendorfer, T., Sickert, S., Ziegler, C.: Rabinizer 4: From {LTL} to your favourite deterministic automaton. In: {CAV} {(1)}. Lecture Notes in Computer Science, vol. 10981, pp. 567--577. Springer (2018)

\bibitem{DBLP:journals/lmcs/KuceraEM06}
Kucera, A., Esparza, J., Mayr, R.: Model checking probabilistic pushdown automata. Log. Methods Comput. Sci.  \textbf{2}(1) (2006)

\bibitem{DBLP:conf/cav/KwiatkowskaNP11}
Kwiatkowska, M.Z., Norman, G., Parker, D.: {PRISM} 4.0: Verification of probabilistic real-time systems. In: {CAV}. Lecture Notes in Computer Science, vol.~6806, pp. 585--591. Springer (2011)

\bibitem{LSAZ22}
Lavaei, A., Soudjani, S., Abate, A., Zamani, M.: Automated verification and synthesis of stochastic hybrid systems: A survey. Automatica  \textbf{146}(12) (2022)

\bibitem{DBLP:conf/aaai/LechnerZCH22}
Lechner, M., \v{Z}ikeli\'{c}, D., Chatterjee, K., Henzinger, T.A.: Stability verification in stochastic control systems via neural network supermartingales. In: {AAAI}. pp. 7326--7336. {AAAI} Press (2022)

\bibitem{NonlinearMangasarian}
Mangasarian, O.L.: Nonlinear Programming. Society for Industrial and Applied Mathematics (1994)

\bibitem{DBLP:conf/podc/MannaP89}
Manna, Z., Pnueli, A.: A hierarchy of temporal properties. In: {PODC}. pp. 377--410. {ACM} (1990)

\bibitem{DBLP:books/daglib/0080029}
Manna, Z., Pnueli, A.: Temporal verification of reactive systems - safety. Springer (1995)

\bibitem{DBLP:journals/peerjpre/MeurerSPCRK0MSR16}
Meurer, A., Smith, C.P., Paprocki, M., Cert{\'{\i}}k, O., Rocklin, M., Kumar, A., Ivanov, S., Moore, J.K., Singh, S., Rathnayake, T., Vig, S., Granger, B.E., Muller, R.P., Bonazzi, F., Gupta, H., Vats, S., Johansson, F., Pedregosa, F., Curry, M.J., Saboo, A., Fernando, I., Kulal, S., Cimrman, R., Scopatz, A.M.: Sympy: Symbolic computing in python. PeerJ Prepr.  \textbf{4},  e2083 (2016)

\bibitem{meyn_tweedie_glynn_2009}
Meyn, S., Tweedie, R.L., Glynn, P.W.: Markov Chains and Stochastic Stability. Cambridge Mathematical Library, Cambridge University Press, 2 edn. (2009)

\bibitem{DBLP:books/daglib/0034861}
Mohri, M., Rostamizadeh, A., Talwalkar, A.: Foundations of Machine Learning. Adaptive computation and machine learning, {MIT} Press (2012)

\bibitem{DBLP:conf/tacas/MouraB08}
de~Moura, L.M., Bj{\o}rner, N.S.: {Z3:} an efficient {SMT} solver. In: {TACAS}. Lecture Notes in Computer Science, vol.~4963, pp. 337--340. Springer (2008)

\bibitem{DBLP:conf/hybrid/MuraliTZ24}
Murali, V., Trivedi, A., Zamani, M.: Closure certificates. In: {HSCC}. pp. 10:1--10:11. {ACM} (2024)

\bibitem{DBLP:conf/aaai/NadaliM0024}
Nadali, A., Murali, V., Trivedi, A., Zamani, M.: Neural closure certificates. In: {AAAI}. pp. 21446--21453. {AAAI} Press (2024)

\bibitem{DBLP:journals/tosem/NguyenKWF14}
Nguyen, T., Kapur, D., Weimer, W., Forrest, S.: {DIG:} {A} dynamic invariant generator for polynomial and array invariants. {ACM} Trans. Softw. Eng. Methodol.  \textbf{23}(4),  30:1--30:30 (2014)

\bibitem{pollard_2001}
Pollard, D.: A User's Guide to Measure Theoretic Probability. Cambridge Series in Statistical and Probabilistic Mathematics, Cambridge University Press (2001)

\bibitem{RobbinsSiegmund1971}
Robbins, H., Siegmund, D.: {A convergence theorem for non negative almost supermartingales and some applications}. Optimizing Methods in Statistics pp. 233--257 (1971)

\bibitem{Safra1988}
Safra, S.: On the complexity of omega-automata. In: {FOCS}. pp. 319--327. {IEEE} Computer Society (1988)

\bibitem{DBLP:conf/sas/SankaranarayananSM04}
Sankaranarayanan, S., Sipma, H.B., Manna, Z.: Constraint-based linear-relations analysis. In: {SAS}. Lecture Notes in Computer Science, vol.~3148, pp. 53--68. Springer (2004)

\bibitem{SA13}
Soudjani, S.E.Z., Abate, A.: Adaptive and sequential gridding for abstraction and verification of stochastic processes. SIAM Journal on Applied Dynamical Systems  \textbf{12}(2),  921--956 (2012)

\bibitem{DBLP:journals/toplas/TakisakaOUH21}
Takisaka, T., Oyabu, Y., Urabe, N., Hasuo, I.: Ranking and repulsing supermartingales for reachability in randomized programs. {ACM} Trans. Program. Lang. Syst.  \textbf{43}(2),  5:1--5:46 (2021)

\bibitem{TA14}
Tkachev, I., Abate, A.: Characterization and computation of infinite horizon specifications over markov processes. Theoretical Computer Science  \textbf{515},  1--18 (2014)

\bibitem{TMKA17}
Tkachev, I., Mereacre, A., Katoen, J.P., Abate, A.: Quantitative model checking of controlled discrete-time markov processes. Information and Computation  \textbf{253}(1),  1--35 (2017)

\bibitem{DBLP:conf/hybrid/TkachevA13}
Tkachev, I., Abate, A.: Formula-free finite abstractions for linear temporal verification of stochastic hybrid systems. In: {HSCC}. pp. 283--292. {ACM} (2013)

\bibitem{DBLP:journals/apal/Vardi91}
Vardi, M.Y.: Verification of concurrent programs: The automata-theoretic framework. Ann. Pure Appl. Log.  \textbf{51}(1-2),  79--98 (1991)

\bibitem{DBLP:conf/tacas/WangZ23}
Wang, Y., Zhu, H.: Verification-guided programmatic controller synthesis. In: {TACAS} {(2)}. Lecture Notes in Computer Science, vol. 13994, pp. 229--250. Springer (2023)

\bibitem{DBLP:conf/fossacs/WinklerGK22}
Winkler, T., Gehnen, C., Katoen, J.: Model checking temporal properties of recursive probabilistic programs. In: FoSSaCS. Lecture Notes in Computer Science, vol. 13242, pp. 449--469. Springer (2022)

\bibitem{DBLP:conf/cav/YangZZTPZ23}
Yang, Z., Zhang, L., Zeng, X., Tang, X., Peng, C., Zeng, Z.: Hybrid controller synthesis for nonlinear systems subject to reach-avoid constraints. In: {CAV} {(1)}. Lecture Notes in Computer Science, vol. 13964, pp. 304--325. Springer (2023)

\bibitem{DBLP:conf/qest/YannakakisE05}
Yannakakis, M., Etessami, K.: Checking {LTL} properties of recursive markov chains. In: {QEST}. pp. 155--165. {IEEE} Computer Society (2005)

\bibitem{DBLP:journals/tac/ZamaniEMAL14}
Zamani, M., Esfahani, P.M., Majumdar, R., Abate, A., Lygeros, J.: Symbolic control of stochastic systems via approximately bisimilar finite abstractions. {IEEE} Trans. Autom. Control.  \textbf{59}(12),  3135--3150 (2014)

\bibitem{DBLP:journals/ejcon/0001SRHH12}
Zhang, L., She, Z., Ratschan, S., Hermanns, H., Hahn, E.M.: Safety verification for probabilistic hybrid systems. Eur. J. Control  \textbf{18}(6),  572--587 (2012)

\end{thebibliography}

\appendix
\section{Detailed Proofs}

\subsection{Proof of \Cref{thm:streett-supermartingale}}\label{app:proof-streett-supermartingales}

We invoke \Cref{thm:robbins-siegmund} with 
$V_t = V (X_t)$, $U_t = \epsilon \cdot \mathbf{1}_{A \setminus B}(X_t)$, $W_t = M \cdot \mathbf{1}_{B}(X_t)$,
and obtain 
\begin{equation}
    \Prob\left(  \sum_{t = 0}^\infty  \mathbf{1}_{A \setminus B}(X_t) < \infty \lor \sum_{t = 0}^\infty  \mathbf{1}_B(X_t) = \infty \right) = 1.
    \label{eqn:partial1}
\end{equation}
Note that, since scaling a series by any finite positive constant 
does not affect its finiteness, we replaced
$ \sum_{t = 0}^\infty \epsilon \cdot \mathbf{1}_{A \setminus B}(X_t) < \infty $ with $\sum_{t = 0}^\infty \mathbf{1}_{A \setminus B}(X_t) < \infty $ and 
$ \sum_{t = 0}^\infty M \cdot \mathbf{1}_B(X_t) = \infty$
with $ \sum_{t = 0}^\infty \mathbf{1}_B(X_t) = \infty$. 
Next, we note that \cref{eqn:partial1} implies that there exists a measure-zero set $N$,
such that every $\omega \in \Omega \setminus N$ satisfies either of the next two cases:
\begin{itemize}
    \item Case $\sum_{t = 0}^\infty \mathbf{1}_B(X_t(\omega)) = \infty$; 
    \item Case $\sum_{t = 0}^\infty \mathbf{1}_B(X_t(\omega)) < \infty \wedge \sum_{t = 0}^{\infty} \mathbf{1}_{A \setminus B}(X_t(\omega)) < \infty$. 
\end{itemize}
We then show that the second case implies that $\sum_{t = 0}^\infty \mathbf{1}_A(X_t(\omega)) < \infty$. To this end, we observe that since $A \subseteq (A \setminus B) \cup B$, the indicator function satisfies $\mathbf{1}_A(\cdot) \leq \mathbf{1}_{A \setminus B}(\cdot) + \mathbf{1}_{B}(\cdot)$ pointwise. 
Therefore, the second case implies that
\begin{align}
    \sum_{t = 0}^{\infty} \mathbf{1}_{A}(X_t(\omega)) &\leq
    \sum_{t = 0}^{\infty} \mathbf{1}_{A \setminus B}(X_t(\omega))
    +\sum_{t = 0}^{\infty} \mathbf{1}_{B}(X_t(\omega)) < \infty\label{eqn:partial2}
\end{align}
This implies that every $\omega \in \Omega \setminus N $ either satisfies 
$\sum_{t = 0}^\infty \mathbf{1}_A(X_t(\omega)) < \infty$ (i.e., \cref{eqn:partial2}) or
$\sum_{t = 0}^\infty \mathbf{1}_B(X_t(\omega)) = \infty$ (i.e., first case), and this establishes \cref{eqn:conclusion-thm2}. \qed

\subsection{Proof of \Cref{thm:requirements-corollary}}\label{app:proof-requirements-corollary}

Firstly, we recall the composition of transition kernel $T^{t}$ defined as 
\begin{equation}\label{eq:kernel}
    T^{t}(a, B) = \int_S T^{t-1}(s, B)~T(a, \mathrm{d}s), \quad 
    T^0(a, B) = \mathbf{1}_{B}(a)
\end{equation}
for every state $a \in S$ and measurable set $B \in \Sigma$. 
Also, we recall that $T^{t}(X_0, B) = \Prob(X_{t} \in B)$ holds for every $B \in \Sigma$.

We then show that \cref{eq:init,eq:invar} imply that, for every $s \in I$, we have $T^{t}(s, I)=1$ for all $t \in \Nat$. We proceed by induction on $t$. 
For the base case, we note that $\forall s \in I \colon T^0(s, I) = \mathbf{1}_I(s) = 1$ by \cref{eq:kernel}.
For the inductive case, we assume the inductive hypothesis 
$\forall s \in I \colon T^{t}(s, I) = 1$ and prove $\forall s \in I \colon T^{t+1}(s, I) = 1$. 
By \cref{eq:kernel} and the inductive hypothesis, for an arbitrary $s \in I$ we have
\begin{equation*}
    T^{t+1}(s, I) = \int_S T^{t}(z, I) ~T(s, \mathrm{d}z)
    \geq
    \int_{I}
    T^{t}(z, I) ~T(s, \mathrm{d}z)
    = 
    \int_{I}
    T(s, \mathrm{d}z)
    = T(s, I)
\end{equation*}
Then, we have that \cref{eq:invar} implies $T^{t+1}(s, I) \geq T(s, I) = 1$, 
which concludes the inductive case. 
Then, by \cref{eq:init} we have $s_0 \in I$ and, since $T^t(s, I) = 1$ for every $s \in I$,
we have $T^t(s_0, I) = 1$. Since $X_0 = s_0$, this establishes that 
$\Prob(X_{t} \in I) = 1$ for all $t \in \Nat$. 
    
    Next, we let $t \in \Nat$ be arbitrary and define $N_t = X_t^{-1}[S \setminus I]$, which has measure zero 
    because $\Prob(X_t \in I) = 1$. 
    We rewrite (iii)--(v) as
    \begin{equation}\label{eqn:cond-3-5-oneline}
        \forall s \in I \colon \Post V(s) \leq V(s) - \epsilon \cdot {\bf 1}_{A\setminus B}(s) + M \cdot {\bf 1}_{B}(s)
    \end{equation}
    We note that, by construction of $N_t$,  
    $\forall \omega \notin N_t \colon X(\omega) \in I$. We thus replace $s$ with $X_t(\omega)$ as rewrite \cref{eqn:cond-3-5-oneline} as follows: 
    \begin{equation*}
        \forall \omega \notin N_t \colon \Post V(X_t(\omega)) \leq V(X_t(\omega)) - \epsilon \cdot {\bf 1}_{A\setminus B}(X_t(\omega)) + M \cdot {\bf 1}_{B}(X_t(\omega))
    \end{equation*}
    By \cref{eqn:postex}, we have that 
    $\forall \omega \notin M_t \colon \Post V(X_t(\omega)) = E(V(X_{t+1}) \mid {\cal F}_t)(\omega)$, for some measure-zero set $M_t$. We define $L_t = N_t \cup M_t$ and rewrite \cref{eqn:cond-3-5-oneline} again: 
    \begin{equation*}\label{eqn:cond-3-5-measure}
        \forall \omega \notin L_t \colon E(V(X_{t+1}) \mid {\cal F}_t)(\omega) \leq V(X_t(\omega)) - \epsilon \cdot {\bf 1}_{A\setminus B}(X_t(\omega)) + M \cdot {\bf 1}_{B}(X_t(\omega))
    \end{equation*}
    Since $L_t$ has measure zero, this establishes \cref{eq:streett-supermartingale} almost surely. By \cref{thm:streett-supermartingale}, we conclude that $V$ is a Streett supermartingale for $(A,B)$. \qed

\subsection{Proof of \Cref{thm:general-streett-pairs}}

By \Cref{thm:streett-supermartingale}, the existence of a Streett supermartingale for each pair ensures that each pair is accepted with probability 1. A finite conjunction of probability 1 events also has probability 1. \qed

\section{Detailed Benchmarks}

\subsection{EvenOrNegative, \Cref{fig:simple}}

\begin{equation*}
    x_{t+1} = \begin{cases}
        2 w_t - 1 &\text{if }x_t = 0\\
        x_t + 1 &\text{if }x_t > 0\\
        x_t - 2 &\text{if }x_t < 0
    \end{cases}
    ,\qquad
    x_0 = 0,\qquad
    w_t \sim \text{Bernoulli}(0.5)
\end{equation*}
\begin{equation*}
    \Phi = {\sf GF}(x \text{ even}) \vee {\sf FG}(x < 0),\quad
    \text{Acc} = \{ (\{ x \in \mathbb{Z} \colon x \geq 0\}, \{x \in \mathbb{Z} \colon x \text{ even} \}) \}
\end{equation*}

\subsection{SafeRWalk1}

\begin{equation*}
    x_{t+1} = x_t + \kappa_0 + (2 w_t-1),
    \quad
    x_0 = 50,\quad
    w_t \sim \text{Bernoulli}(0.5)
\end{equation*}
\begin{equation*}
    \Phi = {\sf G}(x < 100),\qquad
\end{equation*}
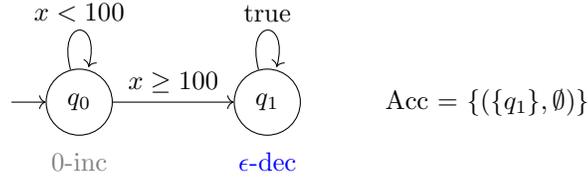
\begin{center}
    \begin{tikzpicture}[initial text={}, node distance=2.5cm]
    \node[state, initial] (Q0) {$q_0$};
    \node[state, right of=Q0] (Q1) {$q_1$};
    \node[below=1mm of Q0] {\color{gray} 0-inc};
    \node[below=1mm of Q1] {\color{blue} $\epsilon$-dec};
    \path (Q0) edge[->] node [above] {$x \geq 100$} (Q1);
    \path (Q0) edge[->, loop above] node [above] {$x < 100$} (Q0);
    \path (Q1) edge[->, loop above] node [above] {true} (Q1);

    \node[right= 1cm of Q1, anchor=west] {\normalsize \text{Acc} = \{$ (
    \{ q_1 \}, \emptyset) \}$};
\end{tikzpicture}
\end{center}

\subsection{PersistRW}
\begin{equation*}
    x_{t+1} = x_t - 0.5 + \frac{1}{10}(2 w_t - 1),\qquad
    x_0 = 50,\qquad
    w_t \sim \text{Bernoulli}(0.5)
\end{equation*}
\begin{equation*}
\Phi = {\sf FG}(x \leq 10),
\end{equation*}
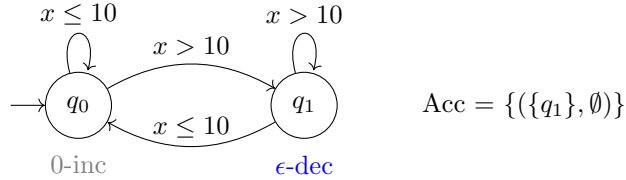
\begin{center}
\begin{tikzpicture}[initial text={}, node distance=3cm]
    \node[state, initial] (Q0) {$q_0$};
    \node[state, right of=Q0] (Q1) {$q_1$};
    \node[below=1mm of Q0] {\color{gray} 0-inc};
    \node[below=1mm of Q1] {\color{blue} $\epsilon$-dec};
    \path (Q0) edge[->, bend left] node [above] {$x > 10$} (Q1);
    \path (Q1) edge[->, bend left] node [above] {$x \leq 10$} (Q0);
    \path (Q0) edge[->, loop above] node [above] {$x \leq 10$} (Q0);
    \path (Q1) edge[->, loop above] node [above] {$x > 10$} (Q1);

    \node[right= 1cm of Q1, anchor=west] {\normalsize \text{Acc} = \{$ (
    \{ q_1 \}, \emptyset) \}$};
\end{tikzpicture}
\end{center}

\subsection{RecurRW}
\begin{equation*}
    x_{t+1} = 
        x_t + 0.1 + 2 (2 w_t - 1)
    ,\qquad 
    x_0 = 50, \qquad
    w_t \sim \text{Bernoulli}(0.5)
\end{equation*}
\begin{equation*}
\Phi = {\sf GF}(x > 100)
\end{equation*}
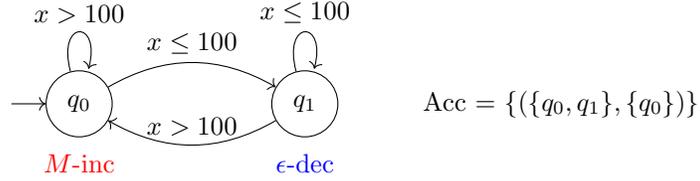
\begin{center}
\begin{tikzpicture}[initial text={}, node distance=3cm]
    \node[state, initial] (Q0) {$q_0$};
    \node[state, right of=Q0] (Q1) {$q_1$};
    \node[below=1mm of Q0] {\color{red} $M$-inc};
    \node[below=1mm of Q1] {\color{blue} $\epsilon$-dec};
    \path (Q0) edge[->, loop above] node [above] {$x > 100 $} (Q0);
    \path (Q0) edge[->, bend left] node [above] {$x \leq 100$} (Q1);
    \path (Q1) edge[->, bend left] node [above] {$x > 100$} (Q0);
    \path (Q1) edge[->, loop above] node [above] {$x \leq 100$} (Q1);

    \node[right= 1cm of Q1, anchor=west] {\normalsize \text{Acc} = \{$ (
    \{ q_0, q_1 \}, \{ q_0 \}) \}$};
\end{tikzpicture}
\end{center}

\subsection{SafeRWalk2}

\begin{equation*}
    x_{t+1} = x_t + \kappa + 10(2w_t - 1),\qquad
    x_0 = 50,\qquad
    w_t \sim \text{Bernoulli}(0.5)
\end{equation*}
\begin{equation*}
\Phi = {\sf G}(x \geq 10)
\end{equation*}
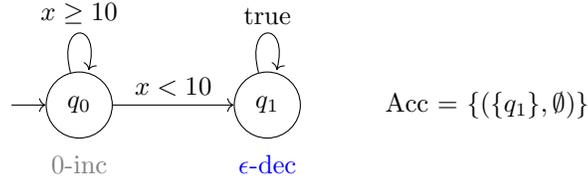
\begin{center}
    \begin{tikzpicture}[initial text={}, node distance=2.5cm]
    \node[state, initial] (Q0) {$q_0$};
    \node[state, right of=Q0] (Q1) {$q_1$};
    \node[below=1mm of Q0] {\color{gray} 0-inc};
    \node[below=1mm of Q1] {\color{blue} $\epsilon$-dec};
    \path (Q0) edge[->, loop above] node [above] {$x \geq 10$} (Q0);
    \path (Q0) edge[->] node [above] {$x < 10$} (Q1);
    \path (Q1) edge[->, loop above] node [above] {true} (Q1);

    \node[right= 1cm of Q1, anchor=west] {\normalsize \text{Acc} = \{$ (
    \{ q_1 \}, \emptyset) \}$};
\end{tikzpicture}
\end{center}

\subsection{GuaranteeRW}
\begin{equation*}
    x_{t+1} = \begin{cases}
        (2 x_t + 1)\cdot 2(2w_t - 1) &\text{if }x_t \geq 0\\
        (2 x_t - 1)\cdot 2(2w_t - 1) &\text{if }x_t < 0
    \end{cases},\quad
    x_0 = 3,\quad
    w_t \sim \text{Bernoulli}(0.5)
\end{equation*}
\begin{equation*}
\Phi = {\sf G}(x \geq -10) \rightarrow {\sf F}(x \geq 10^3),
\end{equation*}
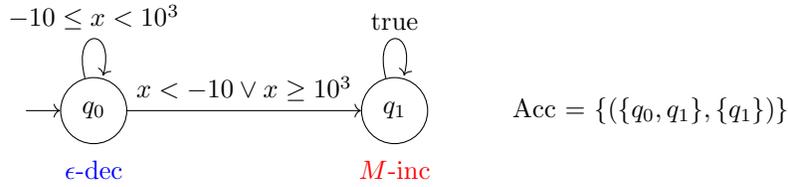
\begin{center}
\begin{tikzpicture}[initial text={}, node distance=4cm]
    \node[state, initial] (Q0) {$q_0$};
    \node[state, right of=Q0] (Q1) {$q_1$};
    \node[below=1mm of Q0] {\color{blue} $\epsilon$-dec};
    \node[below=1mm of Q1] {\color{red} $M$-inc};
    \path (Q0) edge[->, loop above] node [above] {$-10 \leq x < 10^3$} (Q0);
    \path (Q0) edge[->] node [above] {$x < -10 \lor x \geq 10^3$} (Q1);
    \path (Q1) edge[->, loop above] node [above] {true} (Q1);

    \node[right= 1cm of Q1, anchor=west] {\normalsize \text{Acc} = \{$ (
    \{ q_0, q_1 \}, \{ q_1 \}) \}$};
\end{tikzpicture}
\end{center}

\subsection{Temperature1, \cref{eqn:temp-model-dynamics}}
\begin{equation*}
\Phi = {\sf FG}(\neg \text{Hot} \wedge \neg \text{Cold}),
\end{equation*}
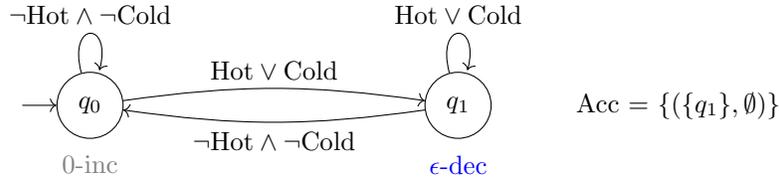
\begin{center}
\centering
\begin{tikzpicture}[initial text={}, node distance=3cm]
    \node[state, initial] (Q0) {$q_0$};
    \node[state, right=4cm of Q0] (Q1) {$q_1$};
    \node[below=1mm of Q0] {\color{gray} 0-inc};
    \node[below=1mm of Q1] {\color{blue} $\epsilon$-dec};
    \path (Q0) edge[->, bend left=8] node [above] {$ \text{Hot} \vee \text{Cold}$} (Q1);
    \path (Q1) edge[->, bend left=8] node [below] {$\neg \text{Hot} \wedge \neg \text{Cold}$} (Q0);
    \path (Q0) edge[->, loop above] node [above] {$\neg \text{Hot} \wedge \neg \text{Cold}$} (Q0);
    \path (Q1) edge[->, loop above] node [above] {$\text{Hot}\vee \text{Cold}$} (Q1);

    \node[right= 1cm of Q1, anchor=west] {\normalsize \text{Acc} = \{$ (
    \{ q_1 \}, \emptyset) \}$};
\end{tikzpicture}
\end{center}

\subsection{Temperature2}
\begin{equation*}
x_{t+1} =
\begin{cases} 
    x_t - 5 + (2w_t - 1) & \text{if } x_t > 40 \\
    x_t - 0.5 + (2w_t - 1) & \text{if } 25 \leq x_t \leq 40 \\
    x_t - 0.1 + (2w_t - 1) & \text{if } x_t < 25 \\
\end{cases},~
x_0 = 35,~
w_t \sim \text{Bernoulli}(0.5)
\end{equation*}
\begin{equation*}
\Phi = {\sf GF}(x \leq 30) \wedge {\sf G}(x \leq 60),
\end{equation*}
\begin{center}
\begin{tikzpicture}[initial text={}]
    \def\edgeshift{1.2cm}
    \node[state, initial] (q0) {$q_0$};
    \node[state,right=3cm of q0] (q1) {$q_1$};
    %\node[state,below=7mm of q1] (q2) {$q_2$};
    \node[state,right=2cm of q1] (q3) {$q_2$};
    \path (q0) edge[bend right=15,->] node[above] {$30 < x \leq 60$} (q1)
    (q1) edge[bend right=15,->] node[above] {$x \leq 30$} (q0)
    (q1) edge[->] node[above] {$x > 60 $} (q3)
    (q1) edge[loop above,->] node[above] {$ 30 < x \leq 60 $} (q1)
    (q0) edge[loop above,->] node[above] {$ x \leq 30 $} (q0)
    (q0) edge[bend right=30,->] node[below] {$x > 60$} (q3)
    (q3) edge[loop above,->] node[above] {true} (q3);
   \node[below=6mm of q3] {$\mbox{Acc} = \{ (\{q_0, q_1, q_2\}, \{ q_0 \})\}$};

    \node[below=0.1cm of q0] {{\color{red} $M$-\text{inc}}};
    \node [below=0.1cm of q1] {{\color{blue} $\epsilon$-\text{dec}}};
    \node [below=0.1cm of q3] {{\color{blue} $\epsilon$-\text{dec}}};
\end{tikzpicture}
\end{center}

\vspace{-4ex}
\subsection{Temperature3 and Temperature4, \cref{eqn:temp-model-dynamics,eqn:shielded-temperature-dynamics}}
\begin{equation*}\label{eqn:temp3-property}
\Phi = {\sf G}(\text{Safe}) \wedge 
    \left[  {\sf GF}(\text{Cold}) \rightarrow 
    {\sf GF}(\text{Hot})\right],\qquad
\end{equation*}
\begin{center}
\begin{tikzpicture}[initial text={}, node distance=3cm]
    \def\edgeshift{1.2cm}
    \node[state, initial] (q0) {$q_0$};
    \node[state,right=3.5cm of q0] (q1) {$q_1$};
    \node[state,right=3.5cm of q1] (q2) {$q_2$};
    \node[state,below=1.5cm of q2] (q3) {$q_3$};

    \path (q0) edge[bend right=15,->] node[above] {$\neg \text{Hot} \land  \text{Cold} \land \text{Safe}$} (q1)
    (q1) edge[bend right=15,->] node[above] {$\text{Hot} \land \text{Safe}$} (q0);

    \path (q1) edge[bend right=15,->] node[above] {$\neg \text{Hot} \land  \neg \text{Cold} \land \text{Safe}$} (q2)
    (q2) edge[bend right=15,->] node[above] {$\neg \text{Hot} \land  \text{Cold} \land \text{S}$} (q1);

    \path (q0) edge[bend right=30,->] node[below] {$\neg \text{Hot} \land \neg \text{Cold} \land \text{Safe}$} (q2)
    (q2) edge[bend right=40,->] node[above] {$\text{Hot} \land \text{Safe}$} (q0);

    \path
    (q3) edge[loop below,->] node[below] {true} (q3);

    \path 
    (q0) edge[bend right=27,->] node[below] {$\neg \text{Safe}$} (q3)
    (q1) edge[bend right=20,->] node[below=1.5mm] {$\neg \text{Safe}$} (q3)
    (q2) edge[->] node[right] {$\neg \text{Safe}$} (q3);

    \path 
    (q0) edge[loop above,->] node[above] {$\text{Hot}\land \text{Safe}$} (q0)
    (q1) edge[loop above,->] node[above] {$\neg \text{Hot} 
    \land \text{Cold} \land \text{Safe}$} (q1)
    (q2) edge[loop above,->] node[above] {$\neg \text{Hot} \land \neg \text{Cold} \land \text{Safe}$} (q2)
    ;

    \node[below=2.7cm of q1] {$\mbox{Acc} = \{ (\{q_1, q_3\}, \{ q_0 \})\}$};

    \node[below=0.1cm of q0] {{\color{red} $M$-\text{inc}}};
    \node[below=0.1cm of q1] {{\color{blue} $\epsilon$-\text{dec}}};
    \node [right=0.1cm of q2] {{\color{gray} $0$-\text{inc}}};
    \node [right=0.1cm of q3] {{\color{blue} $\epsilon$-\text{dec}}};

\end{tikzpicture}
\end{center}

\vspace{-3ex}
\subsection{FinMemoryControl, \cref{eqn:finmemorytemplate}}
\begin{equation*}
\Phi = {\sf GF}(x \leq 0) \rightarrow {\sf GF}(x \geq 100)
\end{equation*}
\begin{center}
\begin{tikzpicture}[initial text={}]
    \def\edgeshift{1.2cm}
    \node[state, initial] (q0) {$q_0$};
    \node[state,right=3cm of q0] (q1) {$q_1$};
    \node[state,right=3cm of q1] (q2) {$q_2$};

    \path (q0) edge[bend right=15,->] node[above] {$0 < x < 100$} (q1)
    (q1) edge[bend right=15,->] node[above] {$x \geq 100$} (q0);

    \path (q1) edge[bend right=15,->] node[above] {$x \leq 0$} (q2)
    (q2) edge[bend right=15,->] node[above] {$0 < x < 100$} (q1);

    \path (q0) edge[bend right=30,->] node[below] {$x \leq 0$} (q2)
    (q2) edge[bend right=40,->] node[above] {$x \geq 100$} (q0);

    \path 
    (q0) edge[loop above,->] node[above] {$x \geq 100$} (q0)
    (q1) edge[loop above,->] node[above] {$0 < x < 100$} (q1)
    (q2) edge[loop above,->] node[above] {$x \leq 0$} (q2)
    ;

    \node[below=6mm of q2] {$\mbox{Acc} = \{ (\{q_2\}, \{ q_0 \})\}$};

    \node[below=0.1cm of q0] {{\color{red} $M$-\text{inc}}};
    \node[below=0.1cm of q2] {{\color{blue} $\epsilon$-\text{dec}}};
    \node [below=0.1cm of q1] {{\color{gray} $0$-\text{inc}}};
\end{tikzpicture}
\end{center}

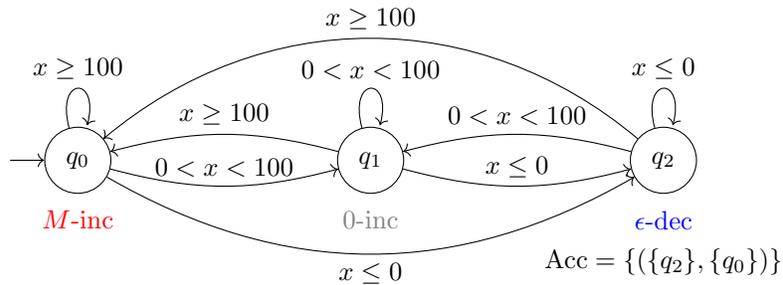
\end{document}